\documentclass[journal]{IEEEtran}
\ifCLASSINFOpdf
\else
\fi

\usepackage{cite}
\usepackage{amsmath,amssymb,amsfonts}
\usepackage{algorithmic}
\usepackage{graphicx}
\usepackage{textcomp}
\usepackage{xcolor}
\usepackage{booktabs}
\usepackage{graphicx}
\usepackage{caption}
\usepackage{subcaption}
\usepackage{siunitx, soul}
\usepackage{adjustbox}
\usepackage{booktabs}
\usepackage{multirow}
\usepackage{hhline}

\usepackage[acronym,toc,nomain]{glossaries}
\newacronym{avc}{AVC}{Advanced Video Coding}
\newacronym{ai}{AI}{All Intra}
\newacronym{alf}{ALF}{Adaptive Loop Filter}
\newacronym{aom}{AOM}{Alliance for Open Media}
\newacronym{amvp}{AMVP}{Advanced Motion Vector Prediction}
\newacronym{avx}{AVX}{Advanced Vector eXtensions}

\newacronym{bt}{BT}{Binary Tree}
\newacronym{bh}{BTH}{Binary Tree Horizontal}
\newacronym{bv}{BTV}{Binary Tree Vertical}
\newacronym{bdr}{BD-BR}{Bj\o ntegaard Delta BitRate}
\newacronym{bdpsnr}{BD-PSNR}{Bj\o ntegaard Delta PSNR}
\newacronym{bdof}{BDOF}{Bi-Directional Optical Flow}

\newacronym{ctu}{CTU}{Coding Tree Unit}
\newacronym{cu}{CU}{Coding Unit}
\newacronym{cnn}{CNN}{Convolution Neural Network}
\newacronym{ctc}{CTC}{Common Test Conditions}
\newacronym{crc}{CRC}{Complexity Reduction Configuration}
\newacronym{cpu}{CPU}{Central Processing Unit}
\newacronym{cabac}{CABAC}{Context Adaptive Binary Arithmetic Coding}
\newacronym{CTC}{CTC}{Common Test Conditions}
\newacronym{ccitt}{CCITT}{International Telegraph and Telephone Consultative Committee}
\newacronym{cb}{CB}{Coding Block}
\newacronym{ciip}{CIIP}{Combined Intra/Inter Prediction}
\newacronym{cc-alf}{CC-ALF}{Cross Component ALF}
\newacronym{cclm}{CCLM}{Cross Component Linear Model}

\newacronym{dl}{DL}{Deep Learning}
\newacronym{DCT}{DCT}{Discrete Cosine Transform}
\newacronym{DST}{DST}{Discrete Sine Transform}
\newacronym{dmvr}{DMVR}{Decoder-side Motion Vector Refinement}
\newacronym{dpb}{DPB}{Decoded Picture Buffer}
\newacronym{dbf}{DBF}{Deblocking Filter}

\newacronym{ecore}{E-core}{Efficient core}

\newacronym{fhd}{FHD}{Full High Definition}
\newacronym{fps}{fps}{Frames Per Second}

\newacronym{gop}{GOP}{Group of Pictures}
\newacronym{gpu}{GPU}{Graphical Processing Unit}
\newacronym{gpm}{GPM}{Geometric Partitioning Mode}

\newacronym{gpp}{GPP}{General Purpose Processor}

\newacronym{hevc}{HEVC}{High Efficiency Video Coding}
\newacronym{hdr}{HDR}{High Dynamic Range}
\newacronym{hi}{HI}{Horizontal Inconsistency}
\newacronym{isp}{ISP}{Intra Sub-Partitionning}
\newacronym{hm}{HM}{HEVC Model}
\newacronym{hd}{HD}{High Definition}
\newacronym{HFR}{HFR}{High Frame Rate}
\newacronym{hvs}{HVS}{Human Visual System}

\newacronym{iso}{ISO}{International Organization for Standardization}
\newacronym{iec}{IEC}{International Electrotechnical Commission}
\newacronym{itu}{ITU}{International Telecommunication Union}
\newacronym{ict}{ICT}{Inter-Component Transform}

\newacronym{jvet}{JVET}{Joint Video Experts Team}
\newacronym{jem}{JEM}{Joint Exploration Model}

\newacronym{qt}{QT}{Quad Tree}
\newacronym{qp}{QP}{Quantization Parameter}
\newacronym{qtbt}{QTBT}{Quad Tree Binary Tree}

\newacronym{lmcs}{LMCS}{Luma Mapping with Chroma Scaling}
\newacronym{ld}{LD}{Low-Delay}
\newacronym{ldp}{LDP}{Low-Delay P}
\newacronym{ldb}{LDB}{Low-Delay B}
\newacronym{lfnst}{LFNST}{Low-Frequency Non-Separable Transform}

\newacronym{mpm}{MPM}{Most Probable Mode}
\newacronym{mv}{MV}{Motion Vector}
\newacronym{ml}{ML}{Machine Learning}
\newacronym{mi}{MI}{Mutual Information}
\newacronym{mdf}{MDF}{Motion Divergence Field}
\newacronym{mc}{MC}{Motion Compensation}
\newacronym{me}{ME}{Motion Estimation}
\newacronym{mpeg}{MPEG}{Moving Picture Experts Group}
\newacronym{mtt}{MTT}{Multi-Type Tree}
\newacronym{MOS}{MOS}{Mean Opinion Score}
\newacronym{MTS}{MTS}{Multiple Transform Selection}
\newacronym{mip}{MIP}{Matrix-based Intra Prediction}
\newacronym{mimd}{MIMD}{Multiple Instructions on Multiple Data}
\newacronym{mse}{MSE}{Mean Squared Error}
\newacronym{mrl}{MRL}{Multiple Reference Line}

\newacronym{pcore}{P-core}{Performance core}
\newacronym{pdf}{Pdf}{Probability Density Function}
\newacronym{pu}{PU}{Prediction Unit}
\newacronym{psnr}{PSNR}{Peak Signal to Noise Ratio}
\newacronym{pmd}{PMD}{Pyramid Motion Divergence}
\newacronym{poc}{POC}{Picture Order Count}
\newacronym{pps}{PPS}{Picture Parameter Set}
\newacronym{prof}{PROF}{Prediction Refinement with Optical Flow}
\newacronym{rd}{RD}{Rate Distorsion}
\newacronym{rdo}{RDO}{Rate Distorsion Optimization}
\newacronym{ra}{RA}{Random Access}
\newacronym{rgb}{RGB}{Red Blue Green}
\newacronym{rf}{RF}{Random Forest}
\newacronym{rs}{RS}{Rectangular Slice}

\newacronym{satd}{SATD}{Sum of Absolute Transform Differences}
\newacronym{svm}{SVM}{Support Vector Machines}
\newacronym{si}{SI}{Spatial Information}
\newacronym{scc}{SCC}{Screen Coding Content}
\newacronym{simd}{SIMD}{Single Instruction on Multiple Data}
\newacronym{shvc}{SHVC}{Scalable High Efficiency Video Coding}
\newacronym{sao}{SAO}{Sample Adaptive Offset}
\newacronym{SPS}{SPS}{Sequence Parameter Set}
\newacronym{sisd}{SISD}{Single Instruction on Single Data}
\newacronym{ssim}{SSIM}{Structural SIMilarity}
\newacronym{sbtmvp}{SbTMVP}{Subblock-based Temporal Motion Vector Prediction}
\newacronym{samviq}{SAMVIQ}{Subjective Assessment Methodology for Video Quality}
\newacronym{sse}{SSE}{Streaming SIMD Extensions}
\newacronym{sota}{SOTA}{State-of-the-Art}

\newacronym{tu}{TU}{Transform Unit}
\newacronym{ti}{TI}{Temporal Information}
\newacronym{tt}{TT}{Ternary Tree}
\newacronym{tth}{TTH}{Ternary Tree Horizontal}
\newacronym{ttv}{TTV}{Ternary Tree Vertical}
\newacronym{trs}{TRS}{Tile and Rectangular Slice}
\newacronym{tl}{TL}{Temporal Layer}
\newacronym{tmvp}{TMVP}{Temporal Motion Vector Prediction}

\newacronym{uhd}{UHD}{Ultra High Definition}

\newacronym{vvc}{VVC}{Versatile Video Coding}
\newacronym{vi}{VI}{Vertical Inconsistency}
\newacronym{vceg}{VCEG}{Video Coding Experts Group}
\newacronym{vtm}{VTM}{VVC Test Model}
\newacronym{vmaf}{VMAF}{Video Multi-Method Assessment Fusion}

\newacronym{wpp}{WPP}{Wavefront Parallel Processing}
\newacronym{waip}{WAIP}{Wide Angular Intra Prediction}

\let\oldtabular=\tabular
\def\tabular{\small\oldtabular}

\newcommand{\Figure}[1] {Fig.~#1}
\newcommand{\Table}[1] {TABLE~#1}

\usepackage{hyperref}
\hypersetup{
    colorlinks=true,
    linkcolor=black,
    filecolor=magenta,      
    urlcolor=black,
}

\begin{document}
	\tolerance 3000
	
	\title{OpenVVC: a Lightweight Software Decoder for the Versatile Video Coding Standard}
	
	%
	%
	
	\author{
		Thomas~Amestoy, 
		Pierre-loup Cabarat, 
		Guillaume Gautier, 
		Wassim Hamidouche,
		 and Daniel Menard
		\thanks{Thomas Amestoy, Pierre-loup Cabarat, Guillaume Gautier, Wassim Hamidouche and Daniel Menard are with Univ. Rennes, INSA Rennes, Institute of Electronic and Telecommunication of Rennes (IETR), CNRS - UMR 6164, VAADER team, 20 Avenue des Buttes de Coesmes, 35708 Rennes, France, (e-mails: \href{mailto:Pierre-Loup.Cabarat@insa-rennes.fr}{Pierre-Loup.Cabarat@insa-rennes.fr}, \href{mailto:wassim.hamidouche@insa-rennes.fr}{wassim.hamidouche@insa-rennes.fr} and \href{mailto:Daniel.Menard@insa-rennes.fr}{Daniel.Menard@insa-rennes.fr}).}
		\thanks{This work was supported by the Energy Efficient Enhanced Media Streaming (3EMS) project. The 3EMS project is jointly funded by Brittany region and Rennes Métropole.}
	}

\markboth{Paper under review, 2022}%
{Shell \MakeLowercase{\textit{et al.}}: Bare Demo of IEEEtran.cls for IEEE Journals}
	
	\maketitle
	
	
	\begin{abstract}
		In the recent years, users requirements for higher resolution, coupled with the apparition of new multimedia applications, have created the need for a new video coding standard.
		The new generation video coding standard, called \gls{vvc}, has been developed by the \acrlong{jvet}, and offers coding capability beyond the previous generation \gls{hevc} standard.
		Due to the incorporation of more advanced and complex tools, the decoding complexity of \gls{vvc} standard compared to \gls{hevc} has approximately doubled.
		This complexity increase raises new research challenges to achieve live software decoding.
		In this context, we developed \emph{OpenVVC}, an open-source software decoder that supports a broad range of \gls{vvc} functionalities.
		This paper presents the \emph{OpenVVC} software architecture, its parallelism strategy as well as a detailed set of experimental results.
		By combining extensive data level parallelism with frame level parallelism, \emph{OpenVVC} achieves real-time decoding of \acrshort{uhd} video content.
		Moreover, the memory required by \emph{OpenVVC} is remarkably low, which presents a great advantage for its integration on embedded platforms with low memory resources. The code of the \emph{OpenVVC} decoder is publicly available at \url{https://github.com/OpenVVC/OpenVVC}.   
	\end{abstract}

	\begin{IEEEkeywords}
		Video compression, VVC, decoding, software, real-time, low memory.
	\end{IEEEkeywords}

	%
	\IEEEpeerreviewmaketitle

	\section{Introduction}\label{section:introduction}
	\IEEEPARstart{D}{uring} the last decade, the extensive use of on-line platforms and the democratization of higher resolutions (4K, 8K) have lead to a significant increase in the volume of exchanged video content~\cite{cisco_global_2021_forecast_highlights_2016}.
	The multimedia services have also diversified with the apparition of video applications that offer immersive and more realistic viewing experience, such as Virtual Reality (VR, 360$^{\circ}$). 
	This increasing demand for video content brings new challenges to compression, mostly to enhance the video coding efficiency and reduce the carbon footprint induced by video storage, transmission and processing.
	Finalized in July 2020,  \gls{vvc}~\cite{hamidouche_versatile_2021, bross_overview_2021} is the \gls{sota} video coding standard. 
	\gls{vvc} has reached the ultimate goal of up to 50\% bit-rate saving compared to \gls{hevc} for similar subjective video quality~\cite{bonnineau2021perceptual, sidaty_compression_2019}. 
	
	The bit-rate savings brought by \gls{vvc} standard over \gls{hevc} are achieved at the expense of more complex coding tools at both encoder and decoder sides.
	The computational complexity of the \gls{vvc} reference encoder has increased by a factor 8 and 27 compared to the \gls{hevc} reference encoder in inter and intra coding configurations, respectively~\cite{bossen_vvc_2021}.
	At the decoder side, which is the focus of this paper, the computational complexity increase of \gls{vvc} standard compared to \gls{hevc} has doubled (2x) in both inter and intra coding configurations~\cite{mercat_comparative_2021}.
	This decoding complexity increase raises new research challenges for \gls{vvc} deployment, especially on embedded platforms or for live applications that require real-time decoding capability.

	Usually, hardware decoders~\cite{6937333} are preferred to software decoders for embedded platforms with low memory and energy supplies.
	However, hardware decoders will only be commercialized several years after the standard finalization.
	The design of efficient software solutions is therefore mandatory during the next couple of years to support real time decoding of the emerging video applications.
	For these applications, the flexibility of software decoders is crucial to support the minor evolution of a standard (ie. new extended profiles), as well as for their deployment on previous generation devices, which do not embed a \gls{vvc} hardware decoder. 
	{Currently only few software decoders compliant with the \gls{vvc} standard have been implemented.
	The \gls{vvc} reference software \gls{vtm}~\cite{noauthor_vvc_2021}, for instance, is compatible with the complete set of new coding tools but requires high memory usage and achieves poor decoding frame rate performance~\cite{mercat_comparative_2021}.
	From the source code of the \gls{vtm}, the Fraunhofer Heinrich Hertz Institute has developed a \gls{vvc} decoder named  \emph{VVdeC}~\cite{vvc_vvdec_2022}.
	This latter offers high decoding speed, at the cost of high memory consumption.}

	In this paper, we present an open source \gls{vvc} software decoder named \emph{OpenVVC}.
	The software architecture, parallelism strategy as well as detailed experimental results are described in this paper.
	\emph{OpenVVC} is developed in C programming language and compiled as a cross-platform library.
	It provides real time decoding capability under various operating systems including MAC OS, Windows, Linux and Android, targeting sustainable real-time decoding of \gls{uhd} content on high performance \gls{gpp} and low performance \gls{gpp} platforms. \emph{OpenVVC} is compatible with popular video players such as FFplay~\cite{noauthor_ffmpeg_2020}, VLC and GPAC.
	Current version of \emph{OpenVVC} decoder supports the decoding of a wide set of conformance videos in addition to the four principal coding configurations defined by \gls{jvet}: \gls{ai}, \gls{ra}, \gls{ldp} and \gls{ldb}.
	
			\begin{figure*}[ht]
		\centering
		\includegraphics[width=0.9\linewidth]{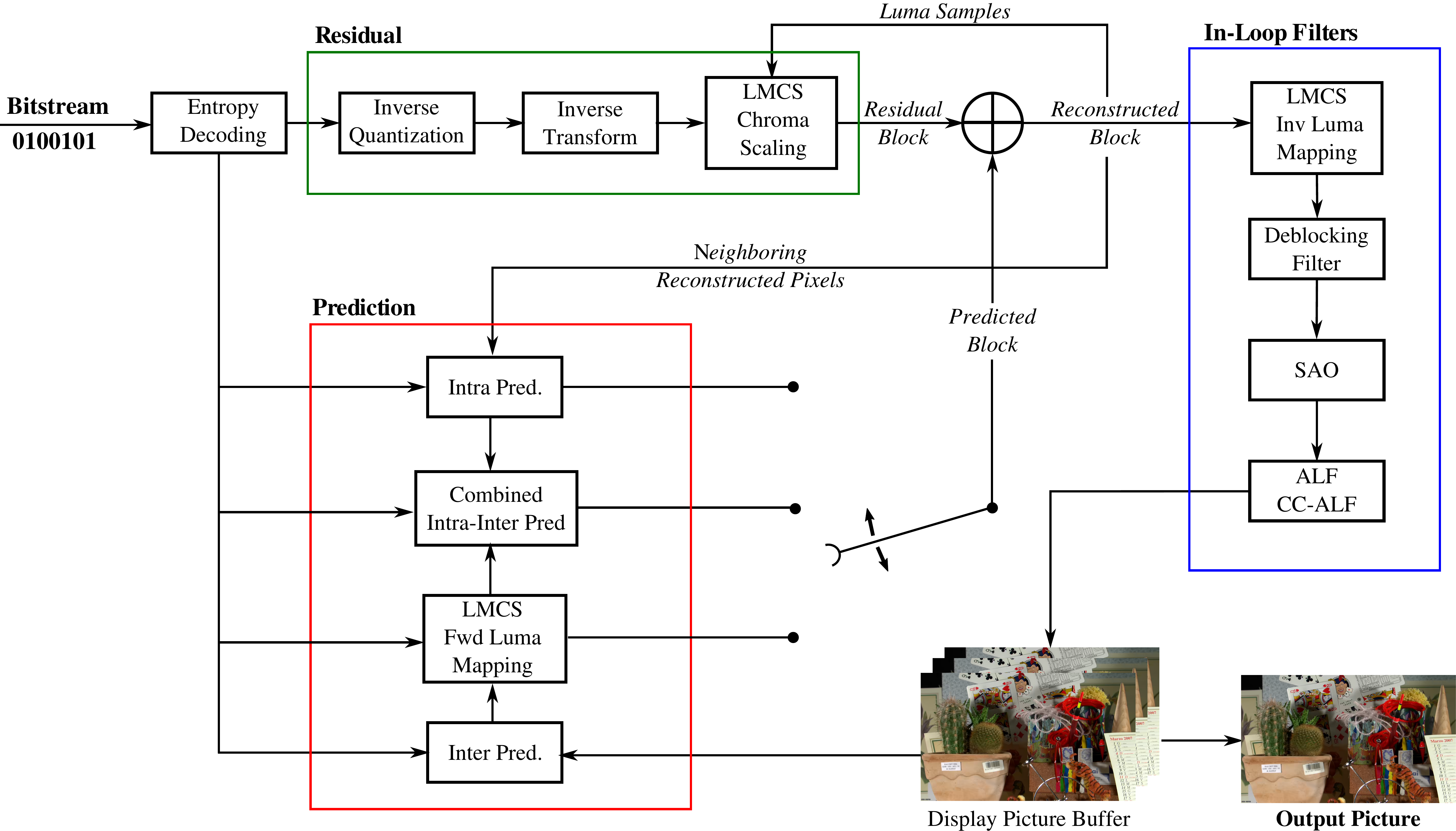}
		\caption{\gls{vvc} decoder block diagram.}
		\label{fig:decoder_vvc}
	\end{figure*}
	
	The \emph{OpenVVC} decoder has been designed to achieve high decoding speed, with the lowest possible memory usage.
	The decoder relies on \gls{simd} optimizations~\cite{bross_hevc_2013} to reduce the decoding time of the most computationally complex operations.
	The architectures of multi-core processor are exploited through frame level parallelism, where several frames are processed simultaneously by the decoder. 
	By combining these two levels of parallelism  on a multi-core x86 platform\footnote{High performance GPP: \url{https://www.intel.fr/content/www/fr/fr/products/sku/134594/intel-core-i712700k-processor-25m-cache-up-to-5-00-ghz/specifications.html}}, \emph{OpenVVC} decoding speed reaches over {290}~\gls{fps} and {90}~\gls{fps} for \gls{fhd} and \gls{uhd} resolutions, respectively.
	Moreover, this high decoding speed is achieved at a very low memory usage.
	The sequential decoding of \gls{fhd} content requires less {than 25~MB and 75~MB in} \gls{ai} and \gls{ra} configurations, respectively. 
	
	The rest of this paper is organized as follows. Section~\ref{section:overview_vvc_dec} presents the general block diagram of a \gls{vvc} decoder and the main decoding stages.
The Section also provides an overview of the \gls{sota} parallelism techniques for the decoding process and presents the \gls{vvc} software decoders currently available. The proposed \emph{OpenVVC} decoder architecture and optimizations are described in detail in Section~\ref{sec:openvvc_presentation}.
	Section~\ref{sec:decoding_results} presents the experimental setup, as well as the performance in terms of memory consumption and decoding speed of \emph{OpenVVC} decoder in \gls{ai} and \gls{ra} coding configurations. In order to highlight the most time consuming tasks of the decoding process, the complexity of \emph{OpenVVC} is also provided by group of tools. Finally, Section~\ref{sec:conc} concludes the paper. 
	
	\section{Background and related work}
	\label{section:overview_vvc_dec}
	In this section we first give the main normative tools integrated in the \gls{vvc} standard and then describe optimizations and parallel processing techniques used to speedup the decoder along with existing software \gls{vvc} decoders.

	\subsection{Overview of VVC decoder}

	\Figure{\ref{fig:decoder_vvc}} presents the general block diagram of a \gls{vvc} decoder, where each block corresponds to one of the main decoding stages.
	The decoder converts an input bitstream composed of binary symbols (bits) into decoded pictures, based on the conventional hybrid coding that takes advantage of both intra/inter prediction and transform coding.
	The operating principles of the decoding stages in \Figure{\ref{fig:decoder_vvc}} are introduced in this section, without going into  in-depth details.
	For a detailed description of the \gls{vvc} tools, the reader may refer to the following papers~\cite{bossen_vvc_2021, hamidouche_versatile_2021}.
	
	\subsubsection{Entropy decoding} 
	
	The first decoding stage is the entropy decoding of the bitstream.
	The \gls{cabac}~\cite{marpe_context-based_2003}, first introduced in \gls{avc} standard, is the entropy engine used in \gls{vvc}.
	At encoder side, the \gls{cabac} has compacted in the bitstream all the syntax elements generated by the coding tools.
	These syntax elements include among others block partitioning information, intra and inter predictions information, quantized coefficients or in-loop filtering parameters.
	At decoder side, the entropy decoding stage parses the  binary symbols in the bitstream, and converts them into non-binary symbols.
	These non-binary symbols are provided as input data for all the other decoding stages.

	\subsubsection{Predicted block}
		
	At the  encoder side, the block partitioning scheme divides the picture into appropriate block sizes according to the local activity of the samples. The block partitioning scheme divides recursively a large block of typical dimension 128$\times$128, called \gls{ctu}, into smaller blocks of sizes in the range 128 $\times$ 64, 64 $\times$ 128 to 4 $\times$ 4 samples, called \glspl{cu}.
	At the decoder side, the \gls{cu} dimensions are retrieved by the entropy decoder and each \gls{cu} is reconstructed by summing a residual block with a predicted block.
	
	The predicted block is an approximation of the original block, computed using intra prediction, inter prediction or a combination of both intra and inter predictions.
	Intra prediction exploits spatial redundancy within the same frame, whereas inter prediction exploits temporal redundancy among adjacent frames. 
	In \gls{vvc}, the novel \gls{ciip} tool combines inter and intra predicted blocks as a weighted sum in order to generate the final predicted block.
	
	When the \gls{cu} is intra predicted, a prediction mode, as well as the previously reconstructed samples of adjacent left and above \glspl{cu} are required.
	These neighboring samples are either propagated with a given angle (angular and \gls{waip} modes), averaged (DC mode), interpolated (Planar mode) or used as input for alternative \gls{mip} mode~\cite{schafer_efficient_2020}.
	The \gls{mrl} mode~\cite{filippov_recent_2019} introduced in \gls{vvc} enables the encoder to choose among three reference lines and explicitly signal the one performing the lowest rate-distortion cost.
	
	For inter prediction, the samples of current \gls{cu} are approximated based on the reference samples stored in the \gls{dpb}.
	This process is called \gls{mc}.
	The decoder first derives one or several \glspl{mv} (whether the \gls{cu} is uni or bi-predicted) from \gls{cabac} information. 
	Each \gls{mv} is composed of a vertical and an horizontal components, representing the underlying samples translation from the reference picture to current picture.
	For bi-predicted blocks, a blending process is applied to aggregate the two motion compensated blocks.
	
	In \gls{vvc}, the \gls{lmcs} has been introduced~\cite{lu_luma_2020} operating in three distinct parts of the decoding process: residual chroma-scaling, forward luma mapping for inter prediction and inverse luma mapping of the reconstructed block.
	For inter prediction, the \gls{lmcs} modifies the luma predicted samples with a forward luma mapping. 
	This process maps (i.e. redistributes) the inter predicted luma samples to another range of values. 
	The residual block being also distributed in the entire possible range of values, this operation is mandatory to sum the inter predicted block with the residual block.
	
	\subsubsection{Residual block}
	
	The inverse quantization and inverse transform are crucial decoding stages in conventional hybrid video codecs.
	The inverse quantization retrieves the value of the transformed residual coefficients, taking as input the quantized coefficients transmitted in the bitstream.
	The transformed residual coefficients are further converted into a non-scaled residual block by the inverse transform.
	The chroma-scaling part of the \gls{lmcs} is finally applied on the non-scaled residual block.
	The scaling factor applied to the chroma samples is computed based on the luma samples of the reconstructed block. The inverse transform module consists in the inverse Low Frequency Non-Separable Transform (LFNST) and the inverse Multiple Transform Selection (MTS).

	\subsubsection{In-loop filters}
	
	Four in-loop filters are performed on the reconstructed samples in order to reduce the visual artifacts of previous coding tools.
	They include the inverse mapping of the \gls{lmcs}, the \gls{dbf}, the \gls{sao} and the \gls{alf}.
	First, the inverse mapping of the \gls{lmcs} redistributes the reconstructed luma samples from the entire possible value range to a smaller range of values.
	Samples in this smaller range of values will be used by all the following in-loop filters and will be stored in the \gls{dpb}.

	The \gls{dbf}~\cite{norkin_hevc_2012} is applied on block boundaries, reducing the blocking artifacts introduced among others by quantization.
	In the decoding process, the horizontal filtering for vertical edges is first performed, followed by vertical filtering for horizontal edges.
	The \gls{sao} ~\cite{fu_sample_2012} then classifies the reconstructed samples into different categories to alleviate the remaining artefacts sample by sample.
	For each category, an offset value, retrieved by the entropy decoder,  is added during the \gls{sao} process to each sample of the category.
	The \gls{sao} is particularly efficient to filter the ringing artefacts and enhance the perceptual video quality.
	
	The last in-loop filters are the \gls{alf}~\cite{tsai_adaptive_2013} and \gls{cc-alf}~\cite{misra_cross_2019}.
	The \gls{alf} performs  Wiener filtering to minimize the \gls{mse} between original and reconstructed samples.
	It is responsible for an important share of \gls{vvc} decoding complexity~\cite{saha_implementation_2016}, mostly due to the classification of every 4$\times$4 block of samples and to the application of diamond shape filters on both luma and chroma samples.
	Applied in parallel with \gls{alf}, the \gls{cc-alf} relies on the luma samples to adjust the chroma samples value.
	
	Once the in-loop filters are performed on the entire picture, the decoded picture is stored in the \gls{dpb}.

	\subsection{Sate-of-the-art of VVC decoding}\label{sec:soa_decoding_par}
	
	The total video decoding workload is determined at the encoder side since it mainly depends on the set of enabled coding tools and on the bitstream final size~\cite{baik_complexity-based_2015}.
	The challenges for a \gls{vvc} decoder are speed and compliance with the standard to support real time decoding of a wide range of bitstreams encoded by different encoders, rates and resolutions. The speed challenge is addressed by parallel processing techniques, which aims to distribute optimally the decoding workload on several actors. 

	In this Section, we first describe the \gls{sota} for parallel decoding techniques.
	Since only few works on \gls{vvc} decoders are currently available, many of the presented related works are designed for \gls{hevc} decoding process.
	It is nonetheless possible to adapt these techniques to a \gls{vvc} decoder, considering that the decoding process of \gls{hevc} and \gls{vvc} standards are quite similar.
	The second part of this section presents the \gls{vvc} software decoders currently available.

	The parallel processing of a decoder essentially operates at three levels of parallelism: data level, high level and frame level.
	The data level parallelism techniques are applied on elementary operations. 
	They include among other techniques relying on \gls{simd} instructions~\cite{bross_hevc_2013}. 
	In high level parallelism techniques, several threads operate on continuous regions of the same frame. 
	These techniques are either normative, i.e. defined in the standard and require additional information in the bitstream, or non-normative.
	With frame level parallelism techniques, several frames are processed in parallel, under the restriction that the \gls{mc} dependencies are satisfied~\cite{hamidouche_4k_2016}.


	\subsubsection{Data level parallelism}\label{subsec:simd}
	
	In the video decoding field, data level parallelism is widely explored through techniques relying on \gls{simd} optimizations~\cite{chi_simd_2015,yan_implementation_2012,hamidouche_4k_2016}.
	With a single \gls{simd} instruction, an operation is applied simultaneously on a vector of data, producing a vector of results.
	For x86  processors, various \gls{simd} set of instructions are available (mainly \gls{sse}~\cite{thakkur_internet_1999} and \gls{avx}~\cite{noauthor_avx2_2011}).
	For instance, the SSE2~\cite{noauthor_sse2_nodate} supports instructions on 128-bit registers, which is 2 and 4 times shorter compared to \gls{avx}2 
	and \gls{avx}512 
	operating on 256-bit and 512-bit registers, respectively.

	The computations that benefit from \gls{simd} architectures are typically those including elementary operations on vectors and matrices.
	In the decoding process, these computations include among others the application of the diamond shape filters of the \gls{alf}, the process of the \gls{mc} interpolation filter, the derivation of reconstructed samples in intra prediction and the inverse transform applied on the residual transform coefficients. 
	On the other hand, the entropy coding stage does not include significant data level parallelism which makes the use of \gls{simd} instructions unnecessary.  
	
	Related works widely rely on \gls{simd} architectures to speed up the decoding process.
	Yan \emph{et al.}~\cite{yan_implementation_2012} rely on intensive \gls{simd} optimizations (SSE instructions on 128-bit registers) to reduce the \gls{hevc} decoding time.
	{The Neon instructions set~\cite{noauthor_arm_nodate}, available on low performance \gls{gpp} processors, is exploited by Raffin \emph{et al.} in their work~\cite{raffin_low_2016}.}
	In the particular case of the scalable extension of \gls{hevc} standard, named \gls{shvc}, Hamidouche \emph{et al.}~\cite{hamidouche_4k_2016} optimize among other the upsampling of the base layer picture with \gls{simd} instructions.
	A complete summary of the possible \gls{simd} optimizations for \gls{hevc} decoding process is provided by Chi {\it  et al.}~\cite{chi_simd_2015}. 
	The authors discuss the challenges of the \gls{simd} implementation for many of the most complex decoding computations, and provide experimental results on 14 different platforms.
	
%

	\subsubsection{Normative high level parallelism}
	
	The normative high level parallelism techniques are defined in the standard and require conveying additional information in the bitstream.
	This subsection presents two of the most widely used normative techniques, namely tiles~\cite{bross_overview_2021} and \gls{wpp}~\cite{wang_high-level_2021}. 
	They have been standardized in both \gls{hevc} and \gls{vvc} in order to facilitate the use of parallel processing architectures for encoding and decoding.
	The proposed \emph{OpenVVC} decoder for instance is compliant with bitstreams including both tiles and \gls{wpp}.

	\textbf{Tiles}

\begin{figure}[ht]
		\centering
		\includegraphics[width=0.65\linewidth]{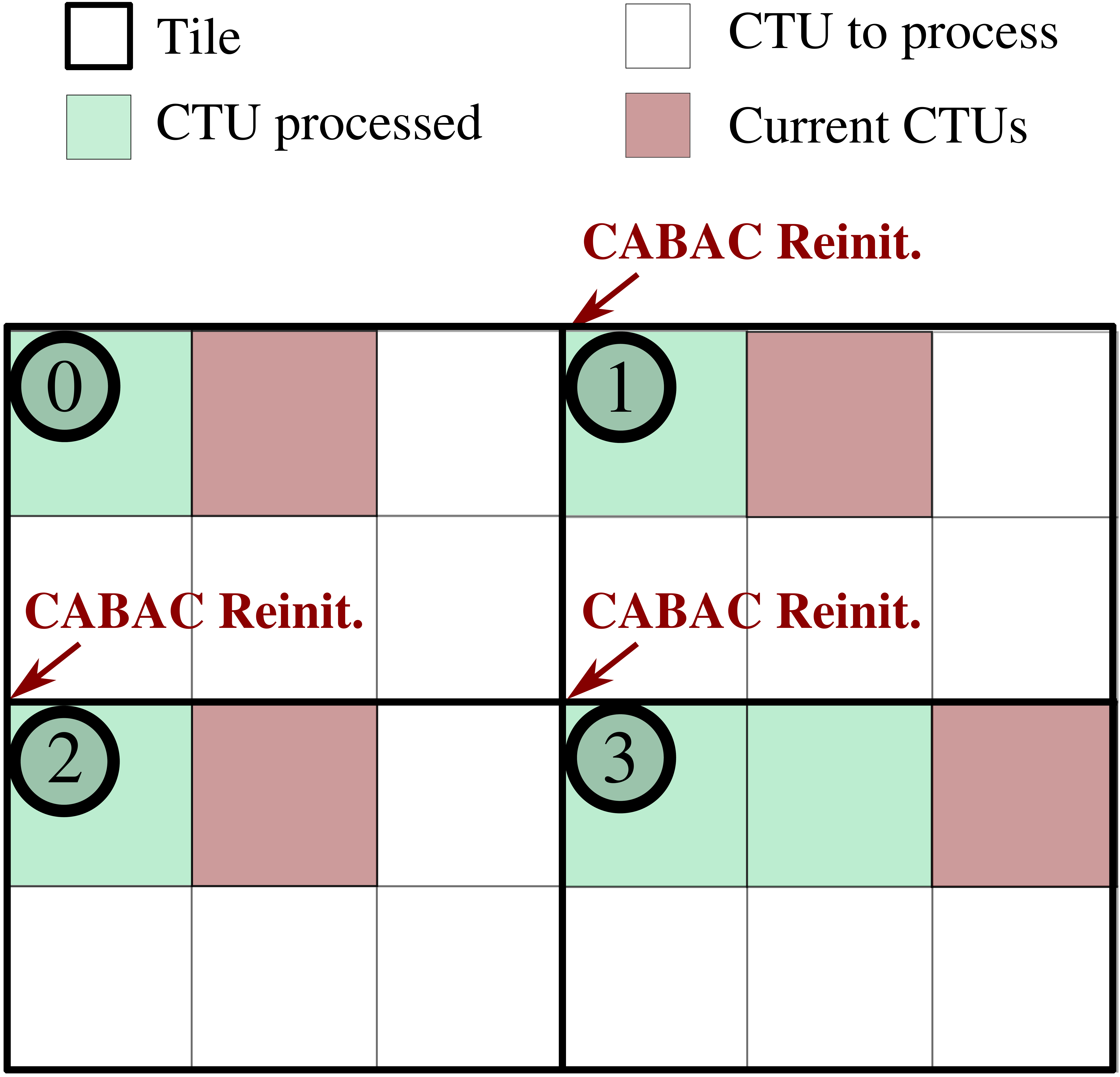}
		\caption{Illustration of tile partitioning: grid of 4 tiles labeled from 0 to 3. The tiles are processed by the decoder in raster scan.}
		\label{fig:hevc_tile}
	\end{figure}

	In both \gls{vvc} and \gls{hevc} standards, tiles are rectangular regions of the picture containing entire \glspl{ctu}~\cite{bross_overview_2021}.
	In \Figure{\ref{fig:hevc_tile}}, the tile partitioning forms a 2$\times$2 grid.
	They are labeled from 0 to 3 and delimited by the thicker black lines.
	Prediction dependencies across tile boundaries are broken and entropy encoding state is reinitialized for each tile.
	These restrictions ensure that tiles are independently decoded, allowing several threads to decode simultaneously the same picture.  
	The in-loop filtering stages across tile boundaries must however be performed when the reconstructed samples of all tiles are available.
	
	In~\cite{baik_complexity-based_2015}, the tile partitioning is adapted at the encoder side in order to minimize the decoding time. 
	The decoding load imbalance among tiles is reduced based on the relation between the decoding time and the number of coded bits of a given \gls{ctu}.
	In the context of computing systems with asymmetric processors, Yoo {\it et al.}~\cite{yoo_parallel_2017} take advantage of tile partitioning flexibility in order to optimize \gls{hevc} decoding time on these specific platforms.
	Asymmetric tile work load is delivered at encoder side by varying tile sizes.
	At the decoder side, the large and small tiles are further allocated to fast and slow cores, respectively.

		\textbf{\acrlong{wpp}}
	
	
	The \gls{wpp} tool, enabled at the encoder side, divides the picture into \gls{ctu} rows~\cite{wang_high-level_2021}.
	The \gls{cabac} context is reinitialized at the start of each \gls{ctu} row with the \gls{cabac} context of at least the second \gls{ctu} in the preceding row. In \gls{vvc}, \gls{wpp} removes the dependency to the top-right \gls{ctu}, thus reducing the \gls{ctu}-offset required between adjacent lines. 
	The decoding of a row may therefore begin once the first \gls{ctu} in the preceding row is reconstructed, since it ensures that the decisions needed for prediction and \gls{cabac} reinitialization are available.
	These constraints allow several processing threads to decode the picture in parallel, with a delay of one \gls{ctu} between adjacent rows.  
	The propagation of these delays across the picture rows limits the parallelism speedup, especially for high number of  threads.
	For this reason, many works including~\cite{chi_improving_2012} and~\cite{shaobo_zhang_implementation_2014} combine \gls{wpp} tool with frame level parallelism in their solutions.
	Zhang {\it et al.}~\cite{shaobo_zhang_implementation_2014} show that the decoder combining frame level parallelism and \gls{wpp} enables much higher speedup compared to the decoder using \gls{wpp} alone, as soon as the number of threads exceeds 5 for high resolution video.

	\subsubsection{Non-normative high level parallelism} \label{subsec:decoding_stage}	
		
	In opposition with decoding parallelism techniques that require specific processing at encoder side, the parallelism techniques presented in this section have not been standardized and are suitable to decode any input bitstream. 
	For instance, task level parallelism techniques refer to techniques in which several threads process simultaneously one or several decoding tasks, exploiting their specific parallel opportunities.
	A detailed description of the main task level parallelism opportunities is provided in~\cite{gudumasu_software-based_2020}.
	Entropy decoding process is sequential and thus it is the most difficult stage to process in parallel.
	In order to process the \gls{cabac} in parallel, \gls{cabac} reinitialization must be included in the bitstream at encoder side, as presented in tiles or \gls{wpp} high level parallelism techniques.
	Habermann \emph{et al.}~\cite{habermann_efficient_2020} propose three solutions to improve the \gls{cabac} processing in \gls{wpp} for low-delay applications. 
	
	Once the \gls{cabac} output data is retrieved, the other decoding tasks may be performed in parallel.
	Two main approaches exist to retrieve the \gls{cabac} output data in related works.
	The first approach performs the \gls{cabac} stage on a picture basis, as a pre-processing of the picture reconstruction.
	This approach, adopted in the \gls{vtm} and in recent works~\cite{gudumasu_software-based_2020, wieckowski_towards_2020}, is optimal for task level parallelism. However, it requires additional memory to store of the \gls{cabac} output data of the whole picture.
	Another approach consists in retrieving the \gls{cabac} information on the fly~\cite{hamidouche_4k_2016}.
	With this approach, task level parallelism is disabled but the memory consumption to store \gls{cabac} output is negligible.
	In this work, the second approach is adopted to lower the memory footprint. 
	The task level parallelism opportunities are therefore not exploited.
	
	For in-loop filtering, a classical approach consists in processing the in-loop filter in a separate pass once the entire picture is reconstructed.
	A slight synchronization overhead is introduced in this case since the in-loop filters are applied one by one on the entire picture.
	Kotra \emph{et al.}~\cite{kotra_comparison_2013} provide three parallel implementations of the \gls{dbf} on the entire reconstructed picture for \gls{hevc} decoding.
	The limit of this approach is that the final samples needed as reference for \gls{mc} are available only after the last in-loop filter is processed on the entire picture.
	However, when in-loop filters are performed on a \gls{ctu} level, the final samples are available at lower delay.
	For instance in~\cite{jo_hybrid_2013}, the final samples of a \gls{ctu} are available with a delay of 2 \glspl{ctu}.
	In the proposed solution, the in-loop filtering is applied at a \gls{ctu} line level. This approach improves the frame level parallelism in inter configuration compared to processing the in-loop filter on the entire picture.

	\subsubsection{Frame level parallelism}\label{subsec:frame_level}
	
	With frame level parallelism, the decoder processes several frames in parallel, under the restriction that the \gls{mc} dependencies are satisfied.
	Frame level parallelism is particularly efficient in \gls{ai} coding configuration since there are no \gls{mc} dependencies.
	In inter coding configuration, its efficiency highly depends on the motion activity in the sequence and on the ranges of \glspl{mv} used for \gls{mc}. 
	Based on this observation, Chi \emph{et al.}~\cite{chi_improving_2012} restrict at the encoder side the downwards motion to 1/4 of image height.
	This restriction reduces greatly the \gls{mc} dependencies for the decoding process, without impacting significantly the coding efficiency.
	
	Frame level parallelism requires a large memory overhead compared to sequential decoding, since the decoder must store additional picture buffer per decoding thread.
	For systems with strong memory constraints such as mobile devices, this memory overhead is a serious limitation.
	For instance in the context of mobile devices, authors in~\cite{rodriguez-sanchez_tiles-and_2017} rely on high level parallelism techniques rather than frame level parallelism alone to accelerate the \gls{hevc} decoding process.

	\subsubsection{VVC software decoders}\label{subsec:software_dec}
	
	Currently only few open-source software \gls{vvc} decoders have been implemented.
	The first to be highlighted is the \gls{vtm}~\cite{noauthor_vvc_2021}, the reference software in which the new tools have been validated during the standardization process.
	Its main advantage is its compatibility with the complete set of new \gls{vvc} standard tools. 
	It has been extensively used during \emph{OpenVVC} development to validate the proper implementation of the new coding tools.
	However, it requires high memory usage and achieves decoding performance far from real time~\cite{mercat_comparative_2021}.
	
	The Fraunhofer Heinrich Hertz Institute has developed an open-source \gls{vvc} decoder named  \emph{VVdeC}~\cite{vvc_vvdec_2022}, from the source code of the \gls{vtm} as a starting point.
	Based on \emph{VVdeC} software, the work of Wieckowski \emph{et al.}~\cite{wieckowski_towards_2020} provides one of the first experimental results on a real-time \gls{vvc} decoder.
	The authors propose intensive \gls{simd} optimizations through \gls{sse}42 (128-bits register) and \gls{avx}2 (256-bits register) instruction sets.
	These optimizations are coupled with task level parallelism, which does not require normative parallel techniques.

	A recent alternative decoder has been proposed by the Tencent Media Lab \emph{O266dec}~\cite{noauthor_tencent_2020}. This decoder is  only available as a binary for testing purpose. 
	Zhu \emph{et al.}~\cite{zhu_real-time_2021} describe the operating principle and experimental results of \emph{O266dec} decoder in \gls{ra} configuration.
	The authors combine \gls{simd} optimization (256-bit register instruction set \gls{avx}2) with techniques exploiting various levels of parallelism: task level, \gls{ctu} level, sub-\gls{ctu} level and frame level.
	The decoding speed for \gls{fhd} and \gls{uhd} video content is very promising.
	However, the task level and \gls{ctu} level parallelism require the storage of the \gls{cabac} information on a picture basis, which may increase the memory consumption.

		\begin{figure}[t]
		\centering
		\includegraphics[width=1\linewidth]{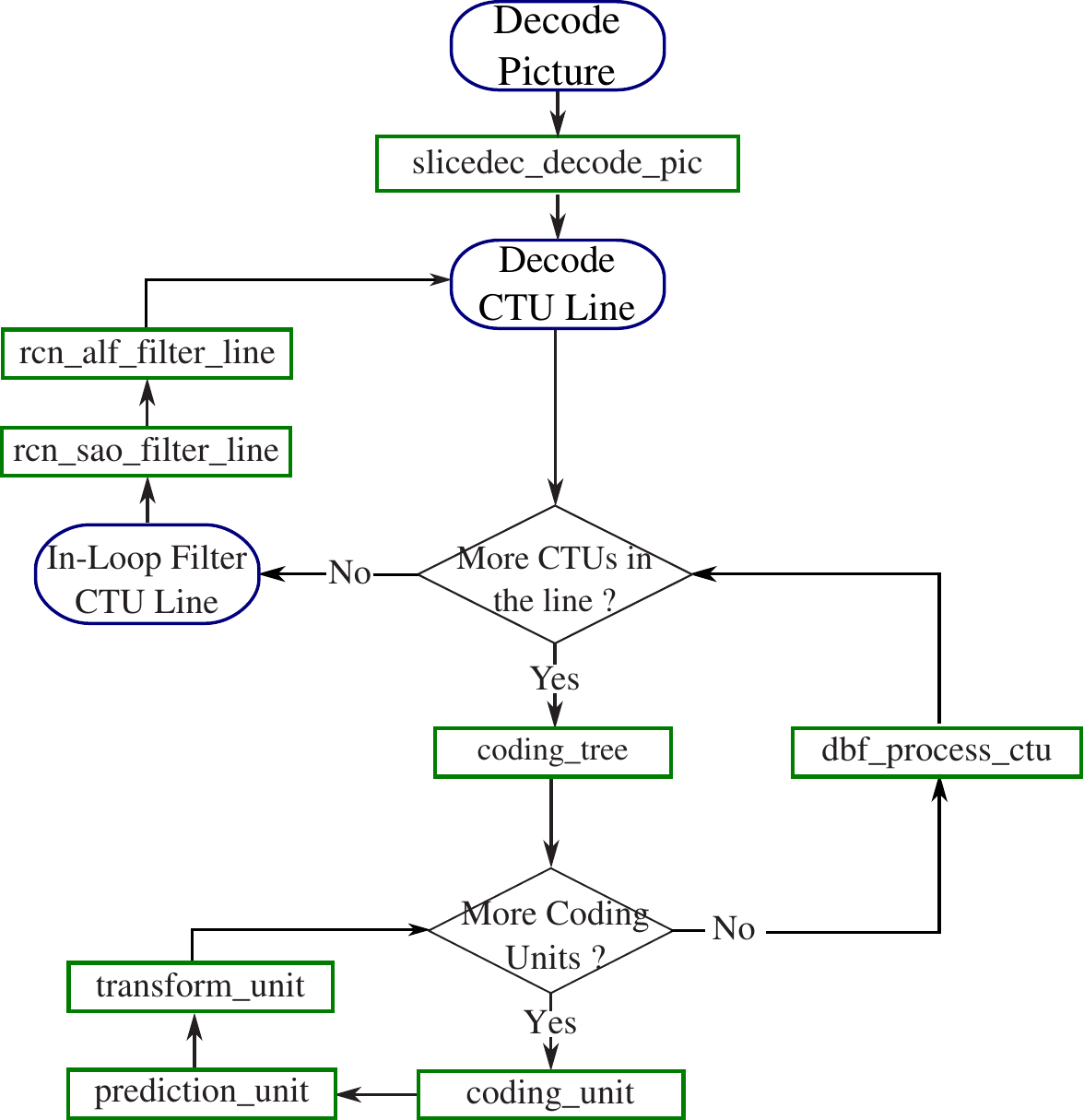}
		\caption{Block diagram of the \emph{OpenVVC} decoder architecture}.
		\label{fig:OpenVVC_diagram}
	\end{figure}
	
	Since 70\% of the world population will have mobile connectivity by 2023 according to Cisco~\cite{cisco_global_2018}, the optimization of the decoding process on low performance \gls{gpp} platforms\footnote{Low performance \gls{gpp}: \url{https://developer.arm.com/tools-and-software/open-source-software/arm-platforms-software}} is a crucial issue. 
	To tackle this concern, Saha \emph{et al.}~\cite{saha_implementation_2016} optimize the \emph{VVdeC} decoder for a system on chip heterogeneous platform composed of a low performance \gls{gpp} and \gls{gpu} processor.
	The \gls{sse} and \gls{avx} instructions included in \emph{VVdeC} are converted to the Neon instruction set available on low performance \glspl{gpp}.
	The results presented do not yet exploit the \gls{gpu} processor, which may decrease the processing time of computationally complex decoding tasks.
	A similar effort was accomplished by Li \emph{et al.}~\cite{li_optimized_2021} to optimize \emph{O266dec} decoder for various low performance \gls{gpp} platforms. The reader can refer to \cite{vvc_implel_2022} for the exhaustive list of available \gls{vvc} encoders and decoders. 
	
	\section {OpenVVC decoder}\label{sec:openvvc_presentation}

	The presented \gls{vvc} decoder is based on the open source \emph{OpenVVC} project.
	The contributors to \emph{OpenVVC} also developed the open source software decoder \emph{OpenHEVC}~\cite{6775908} compliant with \gls{hevc} standard,  used in widespread players such as  VLC\footnote{VLC player: \url{https://www.videolan.org/index.fr.html}} and GPAC~\cite{le_feuvre_gpac_2007}.
	As previously mentioned, the decoder is implemented in C programming language, and is integrated as a dynamic library with FFmpeg player\footnote{OpenVVC library embedded in FFmpeg \url{https://github.com/OpenVVC/OpenVVC}}. 
	The project aims to implement a conforming \gls{vvc} decoder and supports the \gls{ctc}~\cite{boyce_jvet_2018} in \gls{ai}, \gls{ra} and \gls{ld} configuration for 10/8-bits input content.

	This section describes the \emph{OpenVVC} general architecture, its buffer characteristics and the implemented parallelism strategies.

	\subsection{Decoder architecture} \label{subsec:code_struct}

	The decoding parameters required at the sequence, picture, slice or tile level are first retrieved by parsing global parameter sets such as the \gls{SPS}, \gls{pps}, picture header or slice header. 
	The general block diagram for the decoding of a frame in \emph{OpenVVC} is presented in \Figure{\ref{fig:OpenVVC_diagram}}. 
	The main tasks of the decoding process, previously detailed in Section~\ref{section:overview_vvc_dec}, are applied at various levels:
	
	\begin{itemize}
		\item The \gls{cu} level reconstruction includes the intra/inter prediction, inverse quantization, inverse transform and \gls{lmcs} (included in \emph{prediction\_unit} method). 
		
		\item The \gls{dbf} is applied at \gls{ctu} level, right after all the \glspl{cu} of the \gls{ctu} are reconstructed.
		This choice avoids the storage of the \gls{qp} map and \gls{cu} dimensions, required by the \gls{dbf}, on a larger scale.
		
		\item The \gls{sao} filter is applied at a \gls{ctu} line level followed by the \gls{alf}/\gls{cc-alf}.
		This approach improves the frame level parallelism in inter configuration compared to processing the in-loop filters after reconstruction of the entire frame.
	\end{itemize}
	
	\begin{figure}[t]
		\centering
		\includegraphics[width=1.0\linewidth]{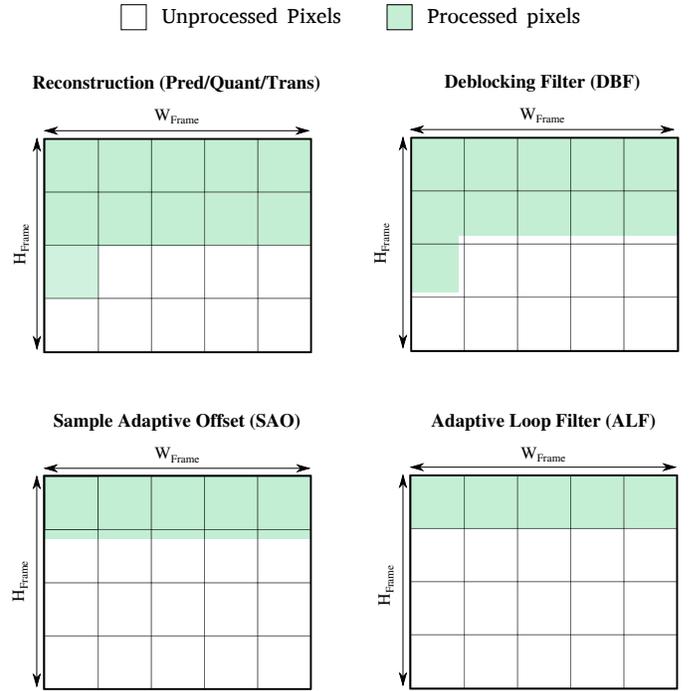}
		\caption{Samples processed by the various decoding stages.}
		\label{fig:decoding_by_task}
	\end{figure}
	
	The decoding process of \emph{OpenVVC} is performed in four successive steps (reconstruction, \gls{dbf}, \gls{sao}, \gls{alf}) as illustrated in \Figure{\ref{fig:decoding_by_task}}.
	The upper-left sub-figure shows the progress of the reconstruction stages : prediction, inverse quantization and transform. In green, the two first \gls{ctu} lines are completely reconstructed.
	As mentioned previously, the \gls{dbf} (upper-right sub-figure) is applied right after the reconstruction on almost all the samples of the current \gls{ctu}.
	A margin of 8 samples is left un-processed at the bottom and at the right of current \gls{ctu}, and will be processed once the required reconstructed samples are available.
	For \gls{sao} filter, a delay of 1 \gls{ctu} line is introduced. 
	The first \gls{ctu} line is entirely processed, as well as a margin of 3 pixel rows in the second \gls{ctu} line.
	This margin is mandatory to apply the \gls{alf} (bottom-right sub-figure) on the entire first \gls{ctu} row.
	The final samples needed as reference for \gls{mc} of other frames are therefore available with a delay of only 1 \gls{ctu} line.

	\subsection{Frame level buffers management} \label{subsec:frame_buffers}

	\begin{figure}[t]
		\centering
		\includegraphics[width=1.0\linewidth]{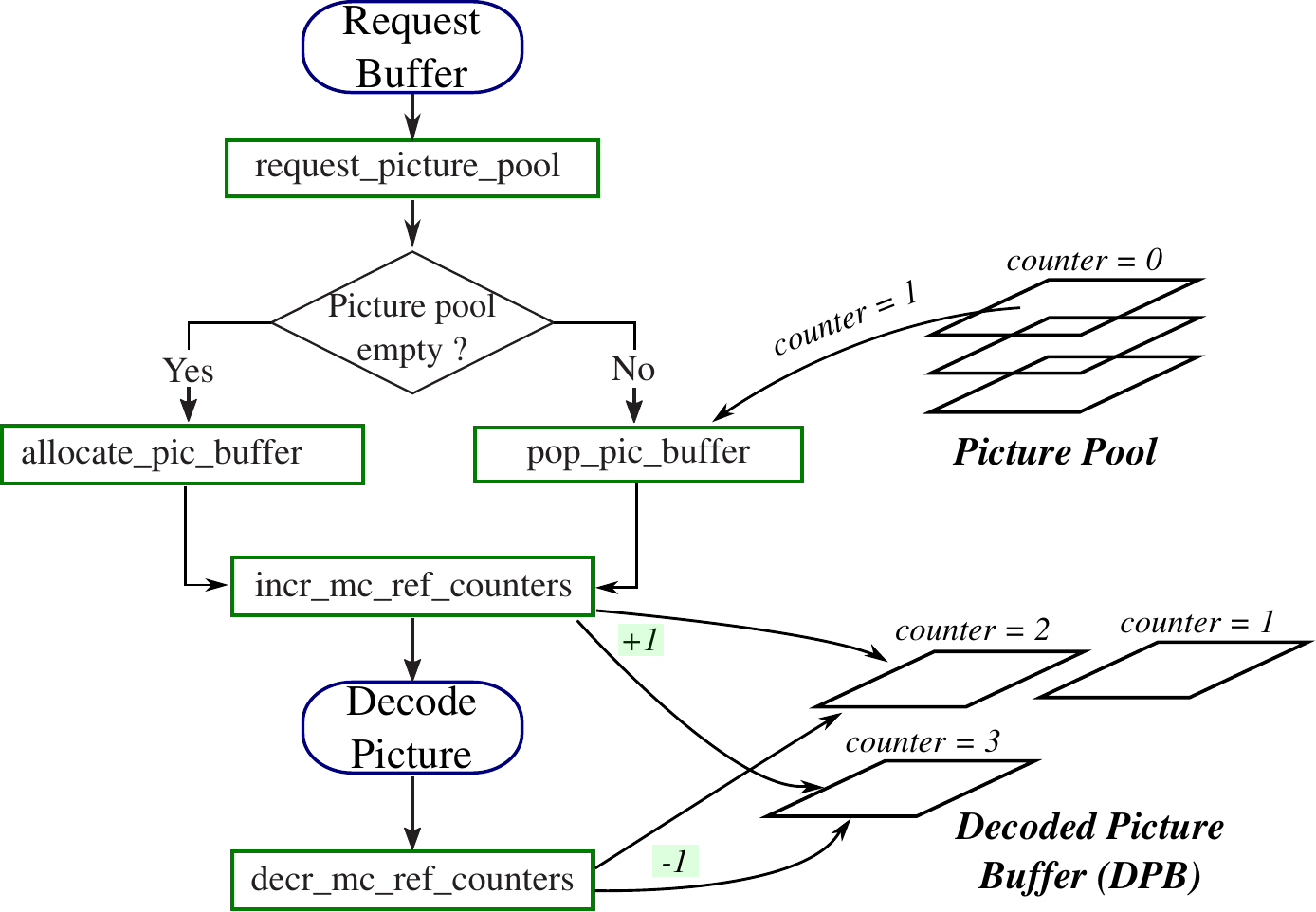}
		\caption{Operating principle of the \gls{dpb} and picture pool in \emph{OpenVVC}}
		\label{fig:dpb_frame_pool}
	\end{figure}
	
	The decoding of a picture in $4:2:0$ chroma format requires a picture buffer with dimensions $1.5 \times W_{frame}\times H_{frame}$, with $W_{frame}$ and $H_{frame}$ the picture width and height, respectively.
	The same picture buffer is used to store all intermediate pixel values during the decoding process.
	The values of the reconstructed samples are overwritten by the \gls{dbf} pixel values, that are further overwritten by the \gls{sao} values and finally replaced by the \gls{alf} output values.
	To ensure that the decoder does not lose intermediate data required for specific stages, local lightweight buffers are used as a complement and are described in next Section~\ref{subsec:local_buffers}.
	
	The picture buffers are counted and a picture pool system manages the unused resources in order to avoid buffer re-allocation (which is time and memory consuming). 
	On the other hand, a \gls{dpb} structure stores the decoded frames that are required for display or used as reference for \gls{mc}.
	\Figure{\ref{fig:dpb_frame_pool}} illustrates the \gls{dpb} management and picture pool in \emph{OpenVVC}. 
	When the decoding of a picture starts, the decoder requests a picture buffer to the picture pool.  
	A new picture buffer is allocated only when the picture pool is empty. 
	Otherwise, the decoder uses a picture buffer popped from the picture pool.
	Then, the decoder increments the counters of the frames in the \gls{dpb} required as reference for \gls{mc} by the current frame. 
	The counters of the reference frames are further decremented when the current picture is decoded.
	When a picture counter falls to 0, it is equivalent to say that it is currently not required neither for display nor as reference for \gls{mc}.
	
	An important aspect for memory consumption limitation, is to remove the unused picture buffers as soon as possible from the \gls{dpb}.
	In the bitstream, an integer \emph{dpb\_max\_nb\_pic} is transmitted, signaling the maximum number of frames needed in the \gls{dpb} for the decoding of a sequence.
	When the current number of frames in the \gls{dpb} exceeds \emph{dpb\_max\_nb\_pic}, the picture with counter equal to 0 and with minimal \gls{poc} is removed from the \gls{dpb} and then is stored in the picture pool.

	The \gls{mv} buffers required for the application of the novel \gls{tmvp} tool are also stored on a frame level.
	These \gls{mv} buffers contain a \gls{mv} for every $8 \times 8$ pixel block of the reference frames.
	A \gls{mv} buffer pool with similar operating principle as the picture buffer pool is implemented in order to avoid buffer re-allocation.

	\subsection{Local-level buffers management} \label{subsec:local_buffers}

\begin{figure}[ht]
		\centering
		\includegraphics[width=0.8\linewidth]{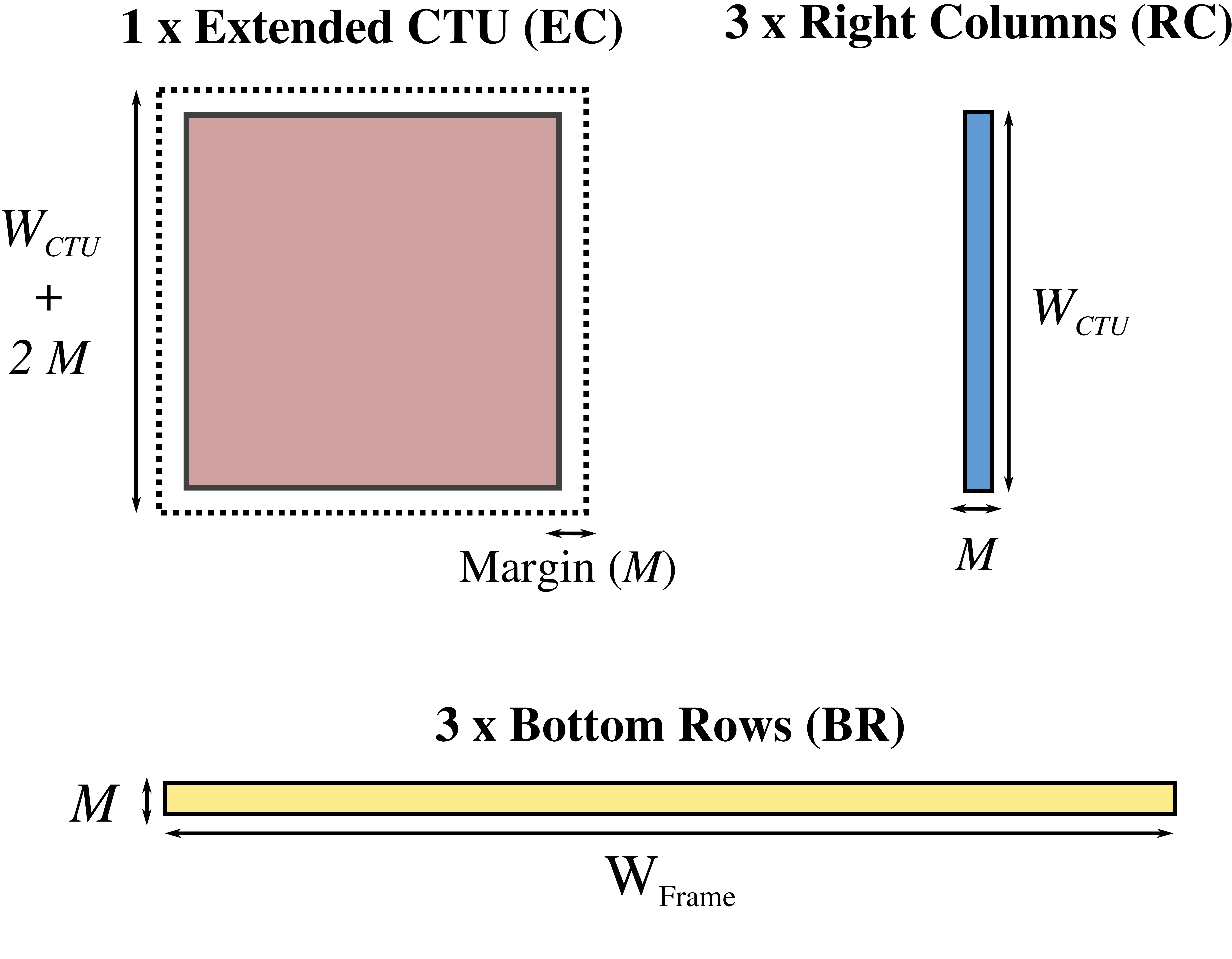}
		\caption{Number and dimensions of local buffers for the decoding on a \gls{ctu} basis.}
		\label{fig:filter_buffers}
	\end{figure}

	The local structure and buffer dimensions have been designed to operate on a \gls{ctu} level.
	As mentioned in Section~\ref{subsec:decoding_stage}, the \gls{cabac} output data required to decode the \gls{ctu} is retrieved on the fly and not stored in the local structure.
	This approach reduces substantially the memory consumption compared to the storage in a frame-basis of the \gls{cabac} output.
	The local structure includes the \gls{ctu} transform residual and \gls{mv} information.
	Moreover, several decoding stages require intermediate samples belonging to neighboring \glspl{ctu}, including intra prediction, \gls{sao} and \gls{alf}.
	The intermediate pixel values are stored in local buffers, and a considerable effort has been made to minimize their dimensions.
	The dimensions of the local buffers are shown in \Figure{\ref{fig:filter_buffers}}, and their usage is further described in this section.

	A single buffer of \gls{ctu} dimensions, called extended \gls{ctu} (EC) buffer, is used to decode the \gls{ctu}.
	The EC buffer is a square block of dimensions $W_{CTU} + 2 \cdot M$, with $M$ the margin of neighboring \gls{ctu} samples.			
	In addition to the EC buffer, each of the 3 decoding stages previously mentioned relies on Right Columns (RC) and Bottom Rows (BR) buffers. The dimensions of the local buffers are shown in \Figure{\ref{fig:filter_buffers}}.
	The BR buffer is a rectangle of dimension $M \times W_{Frame}$ containing the bottom samples of upper \gls{ctu} line.
	The RC buffer is a rectangle of dimension $M \times W_{CTU}$, containing the right samples of the left \gls{ctu}.
	The selected margin $M$ depends on the decoding stage.
	For intra prediction, $M = 3$ since 
	the \gls{mrl} intra prediction mode requires 3 reconstructed pixel columns on the left.
	For in-loop filters, the margin $M$ is half the filter dimension. 
	Therefore, $M = 1$ for the \gls{sao} and $M = 3$ for the \gls{alf}.

	\begin{figure}[ht]
		\begin{minipage}[b]{1.0\linewidth}
			\begin{subfigure}[b]{\linewidth}
				\centerline{\includegraphics[width=0.9\linewidth]{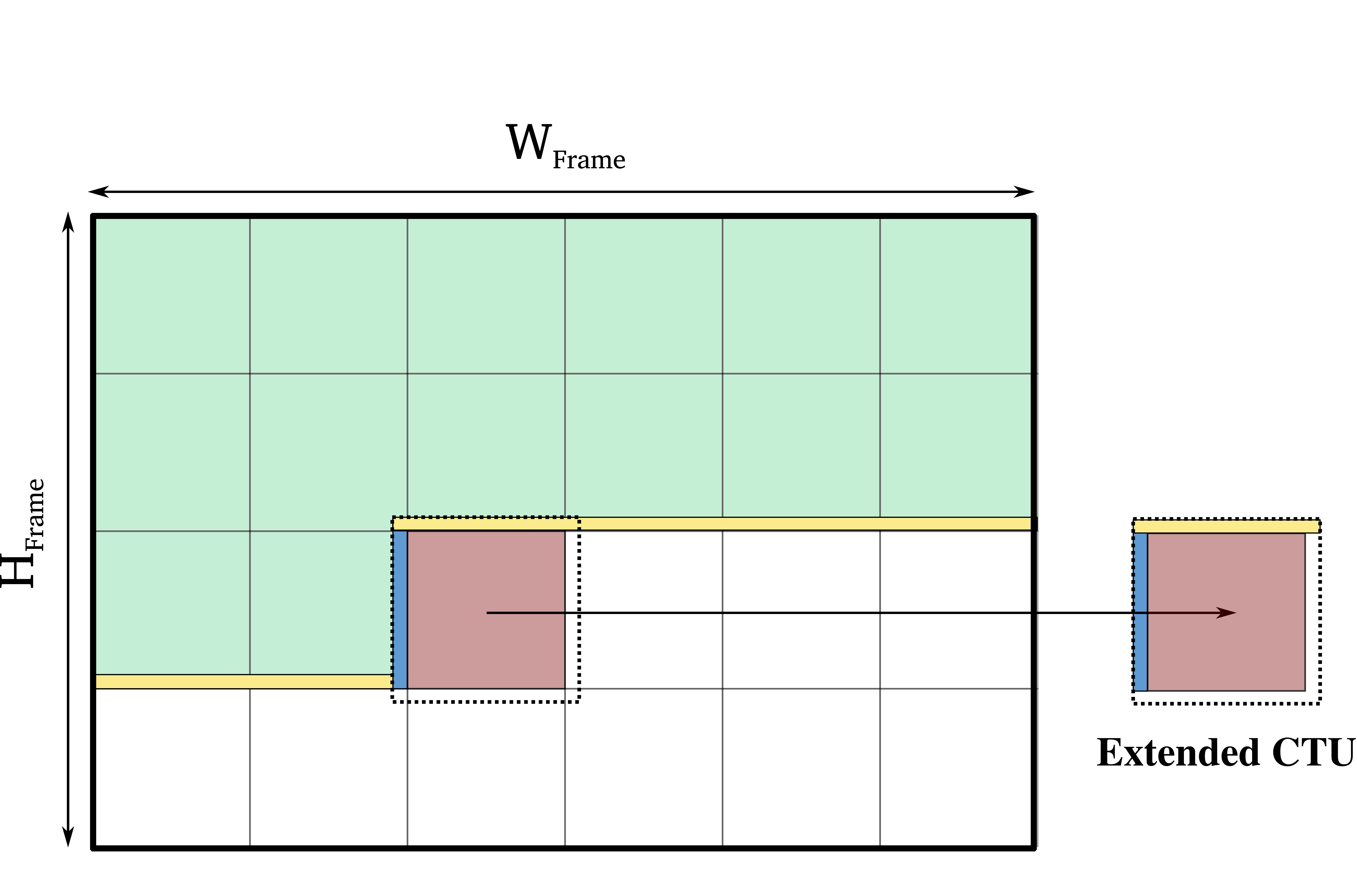}}
				\caption{Fill extended \gls{ctu} buffer with unprocessed samples.}
				\label{fig:filter_usage_fill}
			\end{subfigure}
		\end{minipage} 	
		\begin{minipage}[b]{1.0\linewidth}
			\begin{subfigure}[b]{\linewidth}
				\centerline{\includegraphics[width=0.9\linewidth]{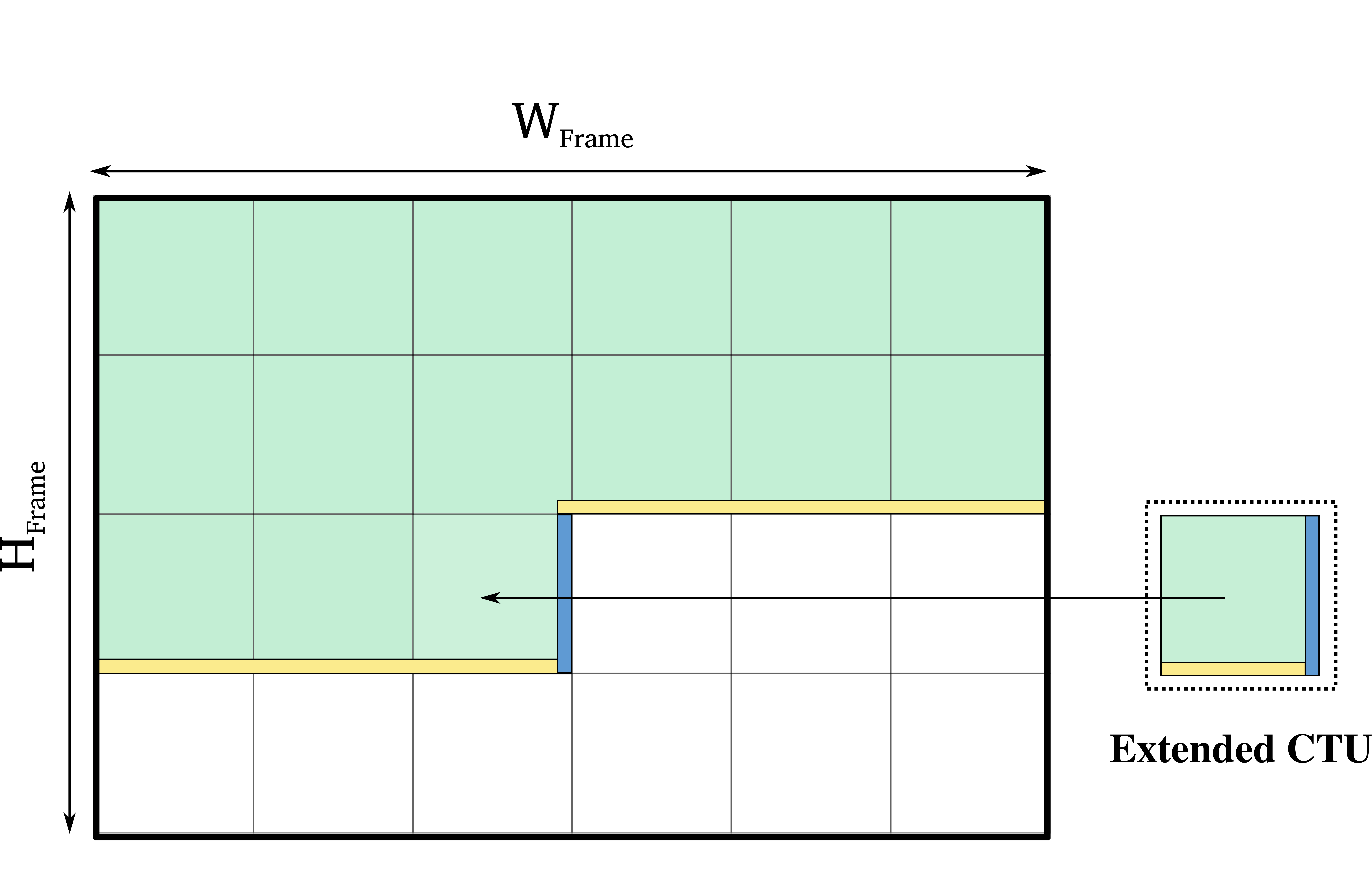}}
				\caption{Update right columns and bottom rows buffers with unprocessed samples and further process \gls{ctu}.}
				\label{fig:filter_usage_apply}
			\end{subfigure}
		\end{minipage}
		\caption{Two steps use of the local buffers on a \gls{ctu} level.}
		\label{fig:filter_usage}
	\end{figure}

	{\Figure{\ref{fig:filter_usage}} illustrates the use of the local buffers for the processing of a given \gls{ctu}. 
	The processes considered in this paragraph are either intra prediction, \gls{sao} or \gls{alf}.
	As shown in \Figure{\ref{fig:filter_usage_fill}}, the EC buffer is first filled with unprocessed samples, i.e. samples on which the process has not been applied.
	The \gls{ctu} area, of dimension $W_{CTU} \times W_{CTU}$, as well as the bottom and right margins are filled directly with the content of the picture buffer.
	The left margin is filled with the unprocessed samples of the left \gls{ctu}, previously stored in the RC buffer.
	The upper margin is filled with the unprocessed samples of the upper \gls{ctu}, previously stored in the BR buffer.
	The second step is shown in \Figure{\ref{fig:filter_usage_apply}}.
	The RC and BR buffers are updated with samples of current \gls{ctu} before processing.
	They will be used during the process of the following \glspl{ctu}.
	Finally, the process is applied on the \gls{ctu} area of the EC buffer that is further copied in the decoded picture.}

	\begin{table}[ht] 
		\centering	
		\caption{Buffers and structures memory consumption for video resolutions FHD and UHD.}
		\label{tb:buffer_sizes}
		\begin{adjustbox}{max width=1\linewidth}
			\begin{tabular}{@{}c|c|c|c|c@{}}
				\hline
				\midrule
				Resolution &  Global Context & Local Context  & Picture Buffer & MV Buffers 
				\\ \midrule
				
				FHD & 5 kB & 741 kB & 8.3 MB & 130 kB
				\\
				
				UHD & 6 kB & 750 kB & 33.2 MB & 620 kB   
				\\ \midrule \hline
				
			\end{tabular}
		\end{adjustbox}
	\end{table}
	
	\Table{\ref{tb:buffer_sizes}} summarizes the memory consumption of the \emph{OpenVVC} buffers and structures, with chroma format 4:2:0, input bit depth 10 and $128\times 128$ \glspl{ctu}.
	The global context structure contains decoding parameters required at the sequence, picture, slice or tile level.
	Since part of the global context parameters are stored on a \gls{ctu} basis, the buffer memory consumption varies slightly from \gls{fhd} to \gls{uhd} sequences.
	The local context structure contains local information required to decode a \gls{ctu}, as well as the local buffers described in Section~\ref{subsec:local_buffers}.
	The local buffers include the BR buffer of dimension $M\times W_{Frame}$, that is larger for \gls{uhd} resolution compared to \gls{fhd}. For this reason the memory consumption is 741KB for \gls{fhd} and 750KB for \gls{uhd}.
	The picture buffer has the dimensions of the picture and is therefore 4 times larger for \gls{uhd} compared to \gls{fhd} resolution.
	The same observation applies to \gls{mv} buffer used for \gls{tmvp}, described in Section~\ref{subsec:frame_buffers}.
	\Table{\ref{tb:buffer_sizes}} shows that the largest share of the memory is consumed by the picture and the local buffers.

	\subsection{Parallelism strategies}\label{subsec:parallelism_strat}

	\emph{OpenVVC} decoder currently supports data level parallelism, frame level parallelism, as well as normative slice level and tile level parallelism (both dynamic and static) defined in the \gls{vvc} standard.
	In this work, we only present the results generated with the combination of data level parallelism and frame level parallelism.
	
	\subsubsection{SIMD optimization}\label{subsec:simd_optim}
	
	\begin{table}[ht]
		\centering
		\renewcommand{\arraystretch}{1.15}
		\caption{Methods optimized with \gls{simd} instructions.}
		\begin{adjustbox}{max width=1\columnwidth}
			\begin{tabular}{l|l}
				\hline
				\midrule
				Module & Method   \\
				\midrule
				\multirow{3}{*}{Transform} 	 & \acrfull{ict}   \\
				&  DST VII, DCT II, DCT VIII   \\
				& \acrfull{lfnst}   \\
				\midrule
				\multirow{5}{*}{Motion Compensation} & Luma 8-tap filters\\
				& Chroma 4-tap filters \\
				& \acrfull{dmvr} \\
				& \acrfull{bdof} \\
				& \acrfull{prof} \\
				\midrule
				\multirow{3}{*}{Intra Prediction}     & DC/Planar \\
				& \acrfull{cclm} \\
				& \acrfull{mip} \\
				\midrule
				\multirow{3}{*}{In-Loop Filters}  & ALF block classification  \\
				& ALF diamond shape filters  \\
				& SAO filter (Edge and Band)  \\
				\midrule
				\hline
			\end{tabular}
		\end{adjustbox}
		\label{tab:simd_methods} 
	\end{table}
	
	For data level parallelism, \emph{OpenVVC} relies on \gls{simd} instruction set \gls{sse}42~\cite{thakkur_internet_1999} operating on 128 bits registers. 
	Several computationally expensive methods are optimized with \gls{simd} instructions, which are summarized in \Table{\ref{tab:simd_methods}}.
	They are mainly distributed into 4 modules~: transform, motion compensation,  intra prediction and in-loop filters.
	Many tools in \Table{\ref{tab:simd_methods}} have been introduced in \gls{vvc} standard, including \gls{ict}, \gls{lfnst}, \gls{cclm},  \gls{mip} and \gls{alf}.
	These tools carry-out computations that benefit from \gls{simd} architectures, since they apply elementary operations at sample level.
	In particular, the \gls{simd} optimization divides by 4 the time consumption of the \gls{alf} diamond shape filters.
	In the future, \emph{OpenVVC} will also rely on \gls{simd} instruction sets with larger registers such as \gls{avx}2 (256-bits) or \gls{avx}512 (512-bits), which will further improve the speed-up offered by data-level parallelism.

	\subsubsection{Frame level parallelism}\label{subsubsec:frame_par}
	
	\begin{figure*}[ht]
		\centering
		\includegraphics[width=1\linewidth]{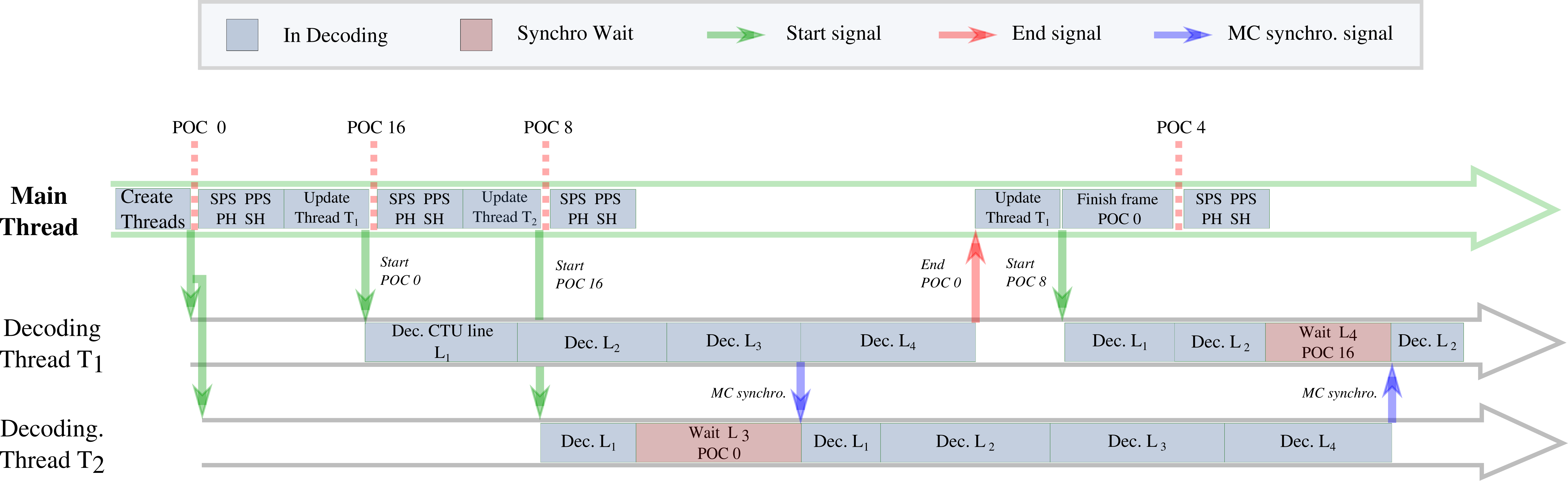}
		\caption{Decoding time-line in \gls{ra} configuration with 2 decoding threads.}
		\label{fig:frame_parallelism}
	\end{figure*}
		
	With frame level parallelism, the decoder processes several frames in parallel, under the restriction that the \gls{mc} dependencies are satisfied.
	Regarding the memory usage, each thread requires separate picture buffer (see Section~\ref{subsec:frame_buffers}) and local buffers (see Section~\ref{subsec:local_buffers}).
	\Figure{\ref{fig:frame_parallelism}} shows an example of a decoding time-line in \gls{ra} configuration with a main thread and 2 decoding threads.
	The main thread is responsible for the parsing of global parameter sets such as the \gls{SPS}, \gls{pps}, picture header or slice header. 
	It also manages the scheduling of the decoding threads through a thread pool.
	Since only frame level parallelism is enabled in this work, the scheduling management of the decoding threads is straightforward. 
	Once the global parameter sets of a picture are parsed, the main thread selects the first available decoding thread in the pool and updates its internal structures with the data required to decode the frame.
	The main thread further signals to the decoding thread (green arrows in \Figure{\ref{fig:frame_parallelism}}), which starts the decoding of the frame.
	When the decoding thread finishes the picture processing, it signals to the main thread (red arrow in \Figure{\ref{fig:frame_parallelism}}) and becomes available again in the thread pool.

	The \gls{mc} synchronization between decoding threads is also a crucial issue for frame level parallelism in inter coding configuration.
	When a decoding thread requires samples not yet available for \gls{mc}, it waits until these samples are reported as available by the thread decoding the reference frame.
	In the example of \Figure{\ref{fig:frame_parallelism}}, the thread $T_2$ decodes \gls{ctu} line $L_1$ in picture $POC_{16}$ and requires the reference data of \gls{ctu} line $L_3$ in picture $POC_0$.
	The thread $T_2$ waits until thread $T_1$ reports available samples in \gls{ctu} line $L_3$ of picture $POC_0$ (blue arrows in \Figure{\ref{fig:frame_parallelism}}), and further continues the decoding of picture $POC_{16}$.
	As explained in Section \ref{subsec:code_struct}, the decoding and in-loop filtering processes are performed on a \gls{ctu} line basis in \emph{OpenVVC}. 
	For this reason, the samples are reported available as reference for \gls{mc} on a \gls{ctu} line basis, once the last in-loop filter is applied. 

	

	\section{Experimental results} \label{sec:decoding_results}
	
	This section presents the experimental setup, as well as the performance in terms of memory usage and frame-rate of the proposed \emph{OpenVVC} decoder in both \gls{ai} and \gls{ra} coding configurations. 
	Data level and frame level parallelism performance is discussed and compared to two open-source \gls{sota} \gls{vvc} decoders: \gls{vtm}-16.2 and \emph{VVdeC}-1.5.
	In order to highlight the most time consuming tasks of the decoding process, the complexity repartition of \emph{OpenVVC} is also provided under the form of pie charts.
	
	\subsection{Experimental setup}
	
	The following experiments are conducted with the proposed \emph{OpenVVC} decoder, in comparison with  the \gls{vtm}-16.2 and \emph{VVdeC} \gls{vvc} decoders. 
	These three open-source software decoders are built with gcc compiler version~11.3.1, under Linux OS version 5.17.6-200 as distributed in Fedora~35.
	The bitstreams decoded during the experiments are generated with the \gls{vtm}-11.2 encoder.
	The platform setup is composed of a 12 cores \gls{gpp}: 8 \glspl{pcore} and 4 \glspl{ecore} \glspl{cpu} running at 3.70~GHz for~\gls{pcore} and 3.60~GHz for~(\gls{ecore}), as detailed in \Table{\ref{tb:platform}}. Each core has 80KB L1 cache~(per core), 1.25MB L2 cache~(per core) and 25MB L3 cache~(shared).
	Moreover, the decoding process is forced to be executed on the \glspl{pcore} and a single set of \gls{simd} instructions is enabled during these experiments (SSE4.2 - 128 bits registers), in order to provide a fair comparison between the software decoders. 
	\begin{table}[th]
		\centering
		\caption{Platform setup used for the decoding performance analysis.}
		\label{tb:platform}
		\begin{adjustbox}{max width=1\columnwidth}
			\begin{tabular}{l|l}
						\midrule
			\midrule
			\textbf{\gls{cpu}}                  & High performance \gls{gpp}\\
			\textbf{Cores}                		& 8 \glspl{pcore} + 4 \glspl{ecore} \\
			\textbf{Clock Rate}                 & 4.70~GHz~(\gls{pcore}) / 3.60~GHz~(\gls{ecore}) \\
			\textbf{SIMD Instructions}          & SSE42 - 128 bits registers \\
			\textbf{Memory}                     & 32 GB \\
			\textbf{Cache (L1, L2, L3)}			& 80~KB / 1.25~MB /  25~MB\\
			\textbf{Compiler}                   & Linux 5.17.6-200-fc35\\
			\textbf{Decoder Version}            & \emph{OpenVVC} v1.1.0 \\
			\textbf{Alternative Decoders}		& VTM-16.2 / \emph{VVdeC}-1.5\\
			\textbf{Encoder Version}          	& VTM-11.2 \\\midrule 			\midrule
			\end{tabular}
		\end{adjustbox}
	\end{table}
	
	The complexity increase of \gls{vvc} decoding process raises a critical issue mainly for high resolution video sequences.
	For this reason, the test sequences selected in this work include 5 \gls{hd} (classes E and F, 1280$\times$720~samples),  6 \gls{fhd} (classes B and F, 1920$\times$1080~samples) and 6 \gls{uhd} (class A, 3840$\times$2160~samples) video sequences included in the \gls{ctc}~\cite{boyce_jvet_2018}.
	Current version of \emph{OpenVVC} is compatible with the encoding tools enabled in the \gls{ctc}, in both \gls{ai} and \gls{ra} configurations.
	The performance of the \emph{OpenVVC} decoder is assessed at various bit-rates, obtained with \gls{qp} values of $ \{ 22,  \,27, \, 32 \, \text{ and } \, 37\}$ following the \gls{ctc}.

	The memory consumption of the software and the output frame-rate are used as performance metrics.
	The maximum memory consumption is measured by { \it time} Linux instruction. It is a crucial information to assess the portability of \emph{OpenVVC} decoder on platforms with strong memory constraints.
	The decoding time is evaluated through the number of decoded frames per second~(fps). 

		\begin{figure*}[ht]
		\begin{minipage}[b]{0.5\linewidth}
			\begin{subfigure}[b]{\linewidth}
				\centerline{\includegraphics[width=0.9\linewidth]{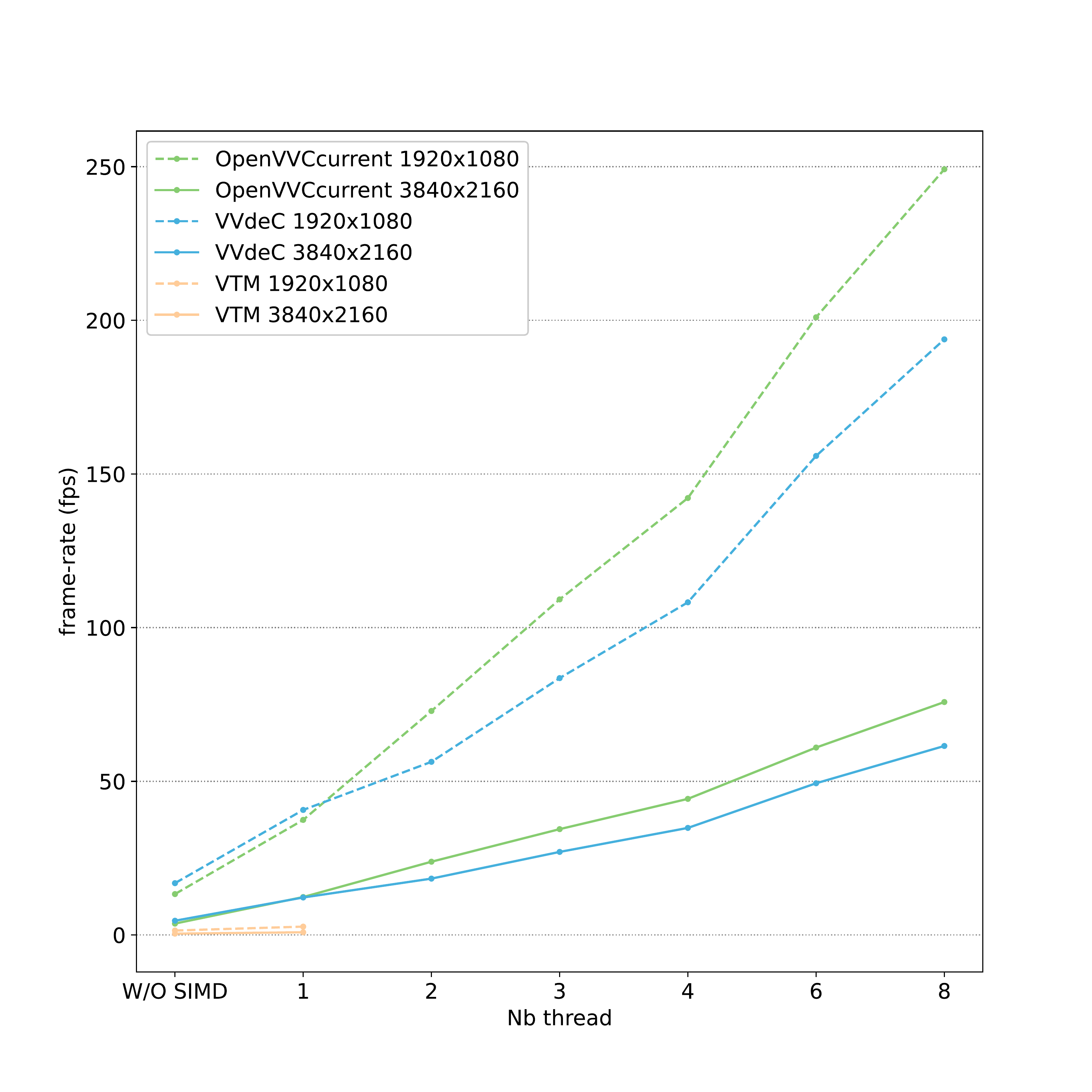}}
				\caption{Decoding frame-rate (in fps) depending on number of threads.}
				\label{fig:compare_fps_ai}
			\end{subfigure}
		\end{minipage} 	
		\begin{minipage}[b]{0.5\linewidth}
			\begin{subfigure}[b]{\linewidth}
				\centerline{\includegraphics[width=0.9\linewidth]{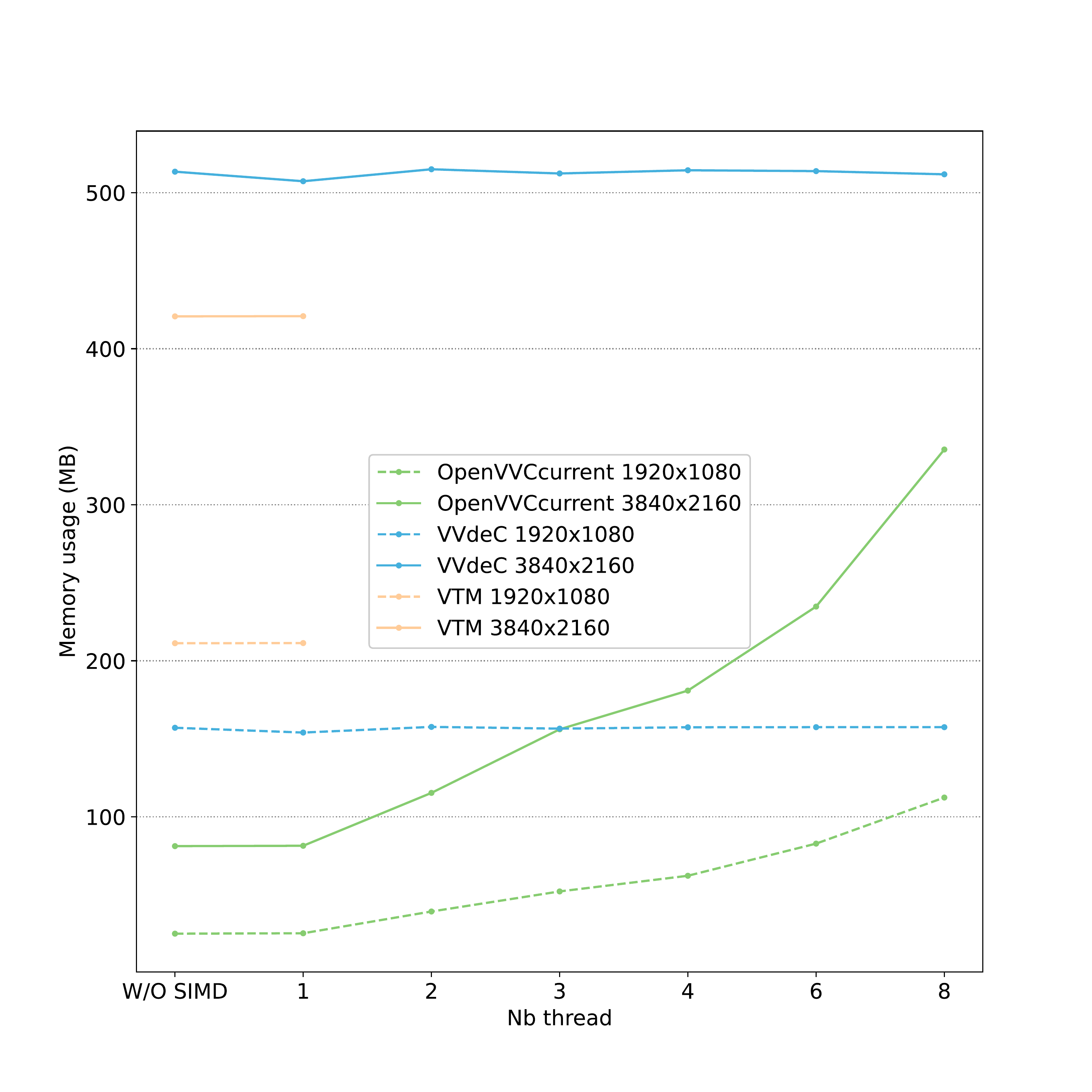}}
				\caption{Memory usage (in KB) depending on number of threads.}
				\label{fig:compare_memory_ai}
			\end{subfigure}
		\end{minipage}
		\caption{\acrlong{ai} configuration : performance of \emph{OpenVVC}, \gls{vtm}-16.2~\cite{noauthor_vvc_2021}  and \emph{VVdeC}~\cite{vvc_vvdec_2022} decoders, averaged across \gls{fhd} and \gls{uhd} test sequences.}
		\label{fig:compare_ai}
	\end{figure*}
	
	\subsection{Comparison with SOTA under AI configuration}\label{subsec:compare_ai}
	
	\Figure{\ref{fig:compare_ai}} presents the performance in \gls{ai} coding configuration of the proposed \emph{OpenVVC} decoder (green points), compared to \gls{vtm}-16.2 (blue points) and \emph{VVdeC} decoders (yellow points), on \gls{fhd} and \gls{uhd} test sequences.
	The two sub-figures correspond to the performance in term of frame-rate (\Figure{\ref{fig:compare_fps_ai}}) and memory consumption (\Figure{\ref{fig:compare_memory_ai}}).
	The results are averaged across all the test sequences with similar resolution and across the 4 \gls{qp} values studied in this work.
	The dashed and continuous lines correspond to \gls{fhd} and \gls{uhd} resolutions, respectively.
	The experiments in this section have been carried out with 7 different parallelism configurations, each corresponding to a different abscissa coordinate.
	These parallel configurations include mono-thread disabling \gls{simd} optimizations, as well as mono, 2, 3, 4, 6 and 8 threads with enabling \gls{simd} optimizations.
	The number of threads does not exceed 8 since software decoders are mainly used on personal computers or smartphones, which rarely exploit architectures with more than 8 cores.
	The \gls{vtm}-16.2 reference software does not support parallel decoding, its performance is therefore assessed through mono-thread setting with disabling and enabling \gls{simd} optimizations.

	\subsubsection{Decoding performance}

	\Figure{\ref{fig:compare_fps_ai}} presents the results in \gls{ai} coding configuration in term of decoding frame-rate, expressed in \gls{fps}.
	First, we will focus on mono-thread results. 
	The points on the left correspond to mono-thread results without \gls{simd} optimizations, and shows almost equivalent decoding speed for \emph{VVdeC} and \emph{OpenVVC}. 
	\Figure{\ref{fig:compare_fps_ai}} shows that the \gls{simd} optimizations are slightly more efficient in \emph{OpenVVC} compared to the two other software decoders.
	This is due to the specific effort dedicated to \gls{simd} optimizations on intra prediction tools, as presented in Section~\ref{subsec:simd_optim}.
	The performance is however still far from real-time since \emph{OpenVVC} achieves an  average frame-rates of {37}~\gls{fps} and {12}~\gls{fps} for \gls{fhd} and \gls{uhd} resolutions, respectively. 
	
	In order to explain multi-thread results, it is important to provide a reminder of the approaches in the different software decoders.
	In \emph{OpenVVC}, frame level parallelism is enabled as presented in Section~\ref{subsec:frame_buffers}.
	Since there are no \gls{mc} dependencies between frames in \gls{ai} configuration, each decoding thread is totally independent.
	For this reason, the frame-rate obtained by \emph{OpenVVC} increases linearly with the number of threads in \Figure{\ref{fig:compare_fps_ai}}, reaching in average {249}~\gls{fps} and {75}~\gls{fps} for \gls{fhd} and \gls{uhd} resolutions, respectively.
	On the other hand, \emph{VVdeC} relies on task level parallelism, introduced in Section~\ref{subsec:decoding_stage}.
	Data dependencies between decoding tasks exist in \gls{ai} configuration, adding synchronization overhead between decoding threads.
	For this reason the frame-rates obtained with more than one thread with \emph{VVdeC} software are lower compared to \emph{OpenVVC}.
	In order to achieve 50 fps decoding of \gls{uhd} resolution, \emph{VVdeC} requires 6 threads in average, while \emph{OpenVVC} only requires 4 threads.

				\begin{figure*}[ht]
		\begin{minipage}[b]{0.5\linewidth}
			\begin{subfigure}[b]{\linewidth}
				\centerline{\includegraphics[width=0.95\linewidth]{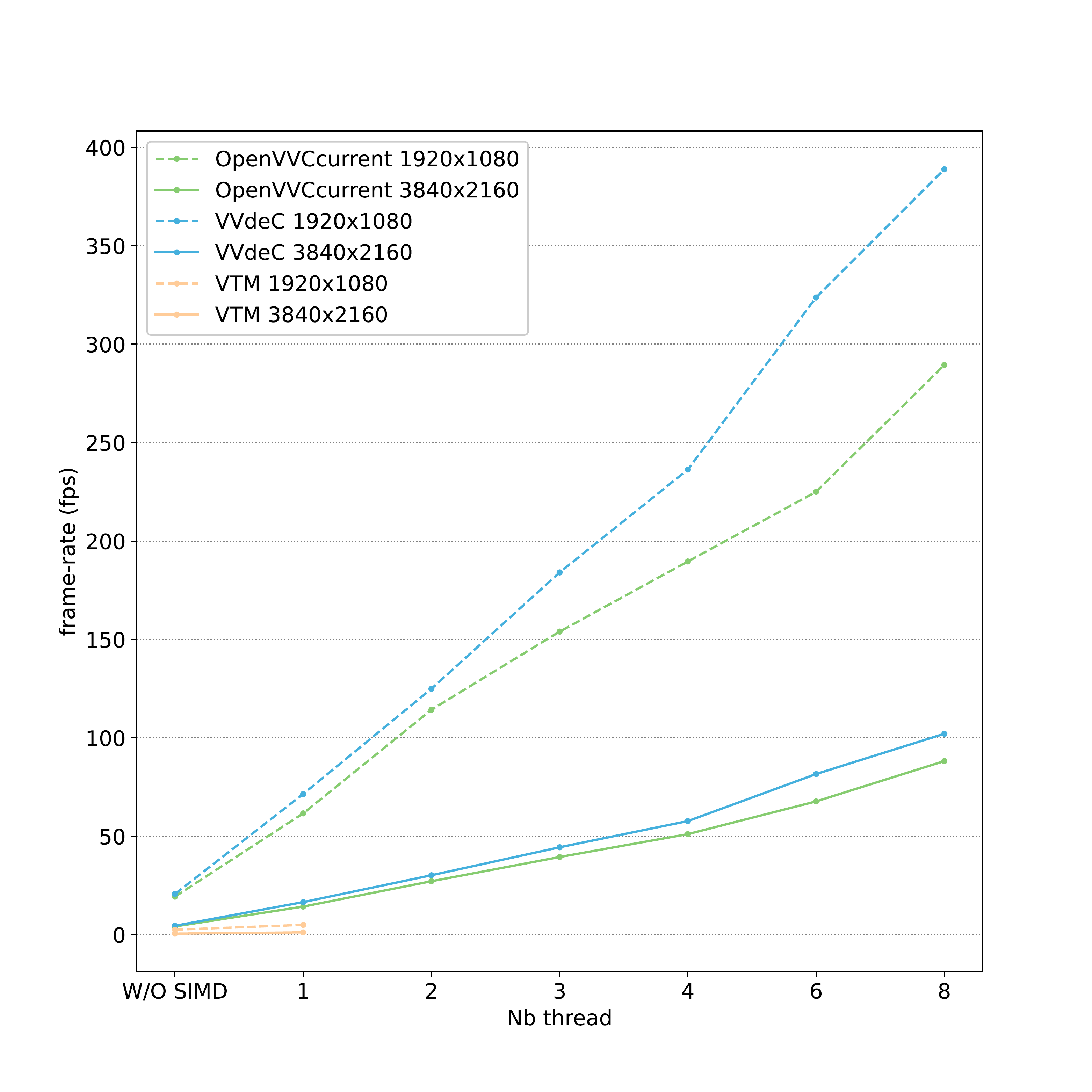}}
				\caption{Decoding frame-rate (in fps) depending on number of threads.}
				\label{fig:compare_fps_ra}
			\end{subfigure}
		\end{minipage} 	
		\begin{minipage}[b]{0.5\linewidth}
			\begin{subfigure}[b]{\linewidth}
				\centerline{\includegraphics[width=0.95\linewidth]{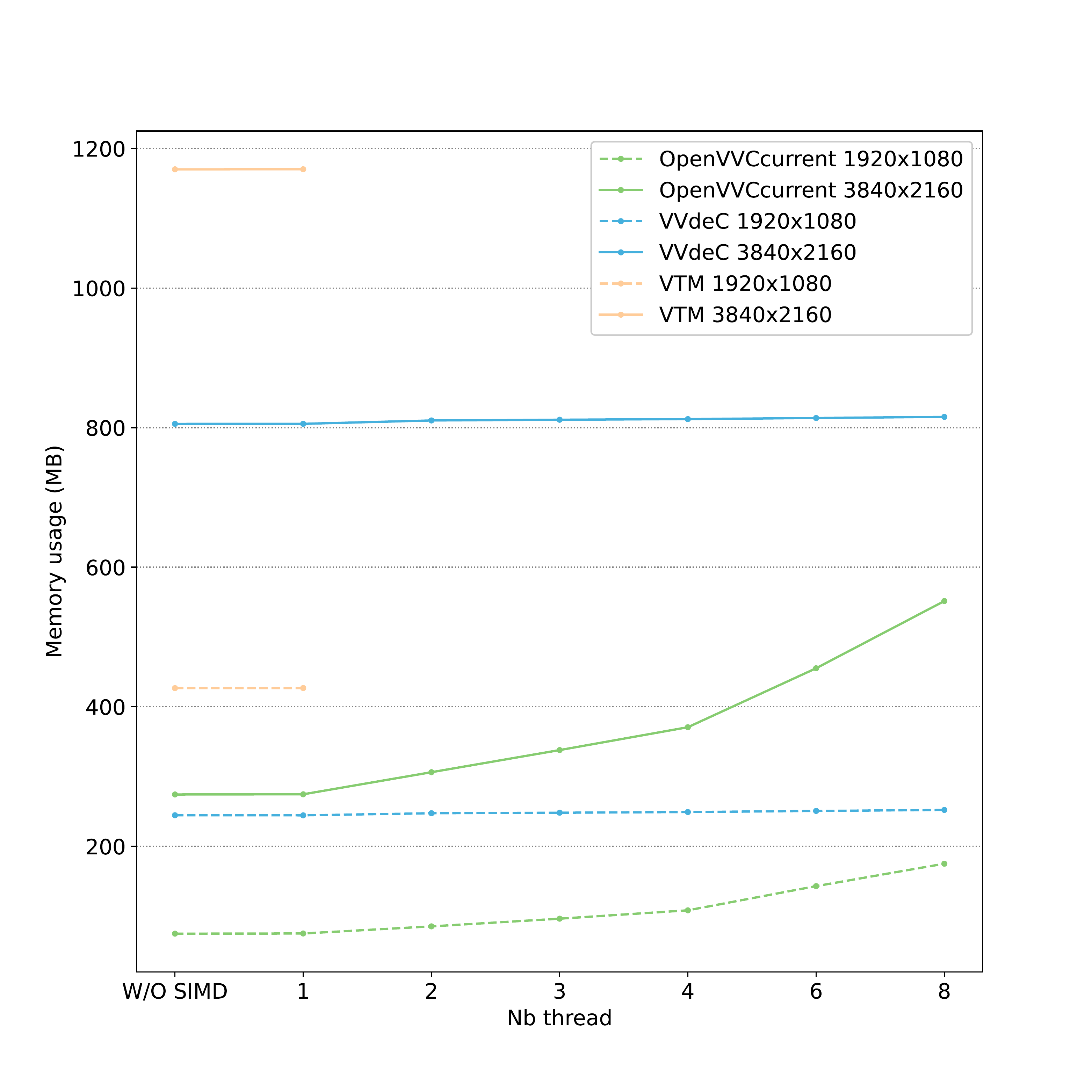}}
				\caption{Memory usage (in KB) depending on number of threads.}
				\label{fig:compare_memory_ra}
			\end{subfigure}
		\end{minipage}
		\caption{\acrlong{ra} configuration : performance of \emph{OpenVVC}, \gls{vtm}-16.2~\cite{noauthor_vvc_2021}  and \emph{VVdeC}~\cite{vvc_vvdec_2022} decoders, averaged across \gls{fhd} and \gls{uhd} test sequences.}
		\label{fig:compare_ra}
	\end{figure*}
	
	\subsubsection{Memory consumption}
	
	\Figure{\ref{fig:compare_memory_ai}} presents the results in \gls{ai} coding configuration in term of memory consumption, expressed in MB.
	With less than 115MB in average, \emph{OpenVVC} is able to decode simultaneously up to 8 frames in \gls{fhd} resolution and up to 2 frames of \gls{uhd} resolution.
	 However, \emph{VVdeC} and \gls{vtm}-16.2 decoders consume 510MB and 420MB, respectively, for the decoding of \gls{uhd} resolution whatever the parallel setting (ie. number of threads).
	These numbers are 1.2 and 1.5 times higher compared to \emph{OpenVVC} with 8 threads. 
	As mentioned in Section~\ref{subsec:decoding_stage}, the memory consumption of these decoders is not optimized. 
	However, it gives an order of magnitude of the very low memory consumption required by the proposed decoding approach in \emph{OpenVVC}.
	This low memory is essentially due to the design of local structure (see Section~\ref{subsec:local_buffers}) and to the optimized management of the picture buffer pool described in Section~\ref{subsec:frame_buffers}.
	
	Second, \Figure{\ref{fig:compare_memory_ai}} shows the linear increase of the memory consumption with the number of threads in \emph{OpenVVC}. 
	As mentioned in Section~\ref{subsec:frame_buffers}, the integer \emph{dpb\_max\_nb\_pic} signals the maximum number of frames required in the \gls{dpb} for the decoding of a sequence. 
	In \gls{ai} configuration, \emph{dpb\_max\_nb\_pic} is equal to 1. 
	For frame level parallelism, a picture buffer and local buffers must be stored in addition per decoding thread, explaining the linear increase.


	\subsection{Comparison with SOTA under RA configuration}
	
	The experiments described in previous Section~\ref{subsec:compare_ai} are also conducted in \gls{ra} coding configuration.
	\Figure{\ref{fig:compare_ra}} presents the performance in \gls{ra} coding configuration of the proposed \emph{OpenVVC} decoder, compared to \gls{vtm}-16.2 and \emph{VVdeC} decoders, on \gls{fhd} and \gls{uhd} test sequences.
	As for \Figure{\ref{fig:compare_ai}}, the results are averaged across the 4 \gls{qp} values studied in this work.
	The dashed and continuous lines correspond to \gls{fhd} and \gls{uhd} resolutions, respectively.

	\subsubsection{Decoding performance}	
	\Figure{\ref{fig:compare_fps_ra}} presents the results in \gls{ra} coding configuration in term of frame-rate, expressed in \gls{fps}.
	\Figure{\ref{fig:compare_fps_ra}} shows that the mono-thread results with \gls{simd} optimizations obtained with \emph{OpenVVC} are higher with almost a factor 2 compared to \gls{vtm}-16.2 for both \gls{fhd} and \gls{uhd} contents.
	On the other hand, \emph{VVdeC} achieves slightly better mono-thread results with \gls{simd} compared to \emph{openVVC}: 12\% and 10\% higher on average for \gls{fhd} and \gls{uhd} resolutions, respectively. 
	In \emph{VVdeC}, significant effort has been invested on data level parallelism, where a larger share of the inter tools are optimized with \gls{simd} instructions compared to \emph{OpenVVC}.
	This small gap will be filled in the future by extending \gls{simd} optimizations to a larger share of the inter prediction tools, adding among other \gls{gpm} and \gls{ciip}.

	\Figure{\ref{fig:compare_fps_ra}} also shows that \emph{OpenVVC} green curves are not completely linear with the number of threads. 
	Indeed, frame level parallelism in \gls{ra} configuration is less efficient compared to \gls{ai} configuration, due to \gls{mc} synchronization overhead between frames.
	\emph{OpenVVC} will achieve higher decoding speed by enabling tile level or task level parallelism in addition to frame level parallelism.
	The results obtained by \emph{OpenVVC} in \gls{ra} configuration are nonetheless very promising, since our decoder achieves live decoding \gls{fhd} sequences beyond 60~fps with in average 2 decoding threads.
	For \gls{uhd} content, picture rate of 50 fps is in average reached with 4 decoding threads.
		
	\subsubsection{Memory consumption}
	\Figure{\ref{fig:compare_memory_ra}} presents the results in \gls{ra} coding configuration in term of memory consumption, expressed in MB.
	In \gls{ra} configuration, the maximum number of frames \emph{dpb\_max\_nb\_pic}  required for the decoding of a sequence is in average equal to 7.
	\Table{\ref{tb:buffer_sizes}} has shown that the picture buffer size is 8.3~MB and 33.2~MB for \gls{fhd} and \gls{uhd} resolutions, respectively.
	This explains the mono-thread memory consumption in \emph{OpenVVC} of 60~MB ($ \approx 7\times 8.3$~MB) and 250~MB ($ \approx 7\times 33.2$~MB) in average for these resolutions.
	For frame level parallelism, a picture buffer and local buffers must be stored in addition per decoding thread, explaining the affine increase of memory consumption with the number of decoding threads.
	
	\Figure{\ref{fig:compare_memory_ra}} also highlights the very low memory consumption of \emph{OpenVVC} in \gls{ra} configuration compared to the \gls{vtm}-16.2 and \emph{VVdeC} software decoders.
	For mono-thread decoding, \emph{OpenVVC} memory consumption is around {30\%} of the\emph{VVdeC} memory in average.
	Even with 8 decoding threads, the memory consumption of our solution represents {65\%} of the\emph{VVdeC} memory for \gls{uhd} resolution.

	\subsection{Per-Sequence performance}

	\begin{table}[t]
		\centering
		\caption{Decoding speed in \gls{ai} configuration according to the sequence, number of threads and \gls{qp} value.}
		\label{tb:sequence_ai}
	\begin{adjustbox}{max width=1\columnwidth}
		\begin{tabular}{c|c|c c c|c c c}
		\hline
					\midrule
		\multicolumn{2}{c|}{} & \multicolumn{6}{c}{FPS}  \\
		\cmidrule{3-8}	
		\multicolumn{2}{c|}{} &  \multicolumn{3}{c|}{QP 27} & \multicolumn{3}{c}{QP 37} \\
		\multicolumn{2}{c|}{} & 1 th. & 4 th. & Sp-up & 1 th. & 4 th. & Sp-up \\
					\midrule
		\multirow{6}{*}{HD} &
		FourPeople              & 75.8	& 288.5	& 3.8	& 113.6	& 441.2	& 3.9 \\
        & Johnny	                & 100.0	& 394.7	& 3.9	& 147.1	& 576.9	& 3.9\\
        & KristenAndSara          & 93.8	& 357.1	& 3.8	& 133.9	& 500.0	& 3.7\\
        & SlideEditing            & 56.7	& 211.1	& 3.7	& 74.5	& 271.4	& 3.6\\
        & SlideShow               & 126.0	& 450.0	& 3.6	& 161.5	& 572.7	& 3.5\\
        & \textbf{Average}        & \textbf{90.4}	& \textbf{340.3}	& \textbf{3.8}	& \textbf{126.1}	& \textbf{472.5}	& \textbf{3.7}\\
					\midrule				
		\multirow{7}{*}{FHD} &
		ArenaOfValor            & 24.9	& 96.2	& 3.9	& 42.6	& 166.7	& 3.9\\
& BasketballDrive         & 35.4	& 134.0	& 3.8	& 58.9	& 217.2	& 3.7\\
& BQTerrace               & 26.4	& 101.4	& 3.8	& 47.2	& 178.6	& 3.8\\
& Cactus	                & 30.4	& 114.5	& 3.8	& 52.5	& 196.9	& 3.8\\
& MarketPlace             & 35.5	& 133.9	& 3.8	& 58.1	& 227.3	& 3.9\\
&  RitualDance             & 39.9	& 150.0	& 3.8	& 63.0	& 241.9	& 3.8\\
& \textbf{Average}        & \textbf{32.1}	& \textbf{121.7}	& \textbf{3.8}	& \textbf{53.7}	& \textbf{204.8}	& \textbf{3.8}\\
					\midrule

		\multirow{7}{*}{UHD} & 
		Campfire                & 9.1	& 33.6	& 3.7	& 18.4	& 65.5	& 3.6\\
& CatRobot1               & 11.2	& 40.9	& 3.6	& 16.7	& 59.4	& 3.5\\
& DaylightRoad2           & 10.7	& 38.4	& 3.6	& 16.2	& 58.5	& 3.6\\
& FoodMarket4             & 13.3	& 47.5	& 3.6	& 18.6	& 64.4	& 3.5\\
& ParkRunning3            & 7.3	& 27.0	& 3.7	& 12.6	& 44.7	& 3.6\\
& Tango2	                & 15.3	& 56.1	& 3.7	& 20.2	& 74.0	& 3.7\\
& \textbf{Average}        & \textbf{11.2}	& \textbf{40.6}	& \textbf{3.6}	& \textbf{17.1}	& \textbf{61.1}	& \textbf{3.6}\\
	
		\midrule
		\hline	
	\end{tabular}
	\end{adjustbox}
	\end{table}

	\Table{\ref{tb:sequence_ai}} presents the decoding speed in \gls{ai} configuration with 1 and 4 threads, according to the video sequence and \gls{qp} value.
	The speed-up obtained with 4 threads is also displayed. 
	For all the sequences, the decoding frame-rate increases with the \gls{qp} value. 
	Indeed, for \gls{qp}37 the bitstream size is in average divided by 2.5 compared to \gls{qp}27.
	It leads a considerable decrease in the decoding complexity due to the lower amount of symbols to process. 
	The relation between the bitstream size and the decoding computational complexity has been studied in detail in~\cite{baik_complexity-based_2015}.
	
	At sequence level, \Table{\ref{tb:sequence_ai}} shows that for a given \gls{qp} value, resolution and number of threads, a high variance in the decoding speed exists according to the sequence characteristics.
	The clearest example is provided by the \gls{hd} sequences, where the frame-rates of \emph{Johnny} and \emph{KristenAndSara} sequences are considerably higher compared to \emph{SlideEditing} video.
	Indeed, \emph{SlideEditing} displays screen content with complex spatial textures.
	It requires a higher number of symbols to be coded, compared to \emph{Johnny} and \emph{KristenAndSara} that display television talk shows with uniform background.
	The decoding speed disparities are also observable among \gls{fhd} sequences, especially between \emph{RitualDance} and \emph{ArenOfValor}, and among \gls{uhd} sequences, especially between \emph{Tango2} and \emph{ParkRunning3}.

	\begin{table}[t]
		\centering
		\caption{Decoding speed in \gls{ra} configuration according to the sequence, number of threads and \gls{qp} value.}
		\label{tb:sequence_ra}
		\begin{adjustbox}{max width=1\columnwidth}

		\begin{tabular}{c|c|c c c|c c c}
		\hline
					\midrule
		\multicolumn{2}{c|}{} & \multicolumn{6}{c}{FPS}  \\
		\cmidrule{3-8}	
		\multicolumn{2}{c|}{} &  \multicolumn{3}{c|}{QP 27} & \multicolumn{3}{c}{QP 37} \\
		\multicolumn{2}{c|}{} & 1 th. & 4 th. & Sp-up & 1 th. & 4 th. & Sp-up \\
					\midrule
		\multirow{6}{*}{HD} &
		FourPeople              & 215.1	& 491.8	& 2.3	& 262.0	& 631.6	& 2.4 \\
        & Johnny	           & 229.9	& 582.5	& 2.5	& 252.1	& 705.9	& 2.8 \\
        & KristenAndSara          & 218.2	& 526.3	& 2.4	& 237.2	& 645.2	& 2.7 \\
        & SlideEditing            & 319.1	& 697.7	& 2.2	& 319.1	& 769.2	& 2.4 \\
        & SlideShow               & 259.1	& 675.7	& 2.6	& 266.0	& 746.3	& 2.8 \\
        & \textbf{Average}        & \textbf{248.3}	& \textbf{594.8}	& \textbf{2.4}	& \textbf{267.3}	& \textbf{699.6}	& \textbf{2.6} \\
					\midrule				
		\multirow{7}{*}{FHD} &
		ArenaOfValor            & 64.4	& 179.6	& 2.8	& 87.6	& 262.0	& 3.0 \\
& BasketballDrive         & 52.0	& 162.3	& 3.1	& 62.8	& 209.2	& 3.3 \\
& BQTerrace               &  55.9	& 170.0	& 3.0	& 68.4	& 218.2	& 3.2 \\
& Cactus	                & 66.0	& 201.6	& 3.1	& 88.5	& 277.8	& 3.1 \\
& MarketPlace             &  51.5	& 160.9	& 3.1	& 67.8	& 219.8	& 3.2 \\
&  RitualDance             &  59.4	& 184.0	& 3.1	& 75.8	& 246.9	& 3.3 \\
& \textbf{Average}        &  \textbf{58.2}	& \textbf{176.4}	& \textbf{3.0}	& \textbf{75.1}	& \textbf{239.0}	& \textbf{3.2} \\
					\midrule

		\multirow{7}{*}{UHD} & 
		Campfire                & 15.8	& 54.9	& 3.5	& 19.5	& 69.0	& 3.5 \\
& CatRobot1               &  14.1	& 50.7	& 3.6	& 17.4	& 62.0	& 3.6 \\
& DaylightRoad2           &  13.2	& 46.8	& 3.6	& 16.2	& 57.5	& 3.5 \\
& FoodMarket4             & 14.0	& 50.4	& 3.6	& 16.5	& 59.4	& 3.6 \\
& ParkRunning3            & 9.4	& 33.0	& 3.5	& 12.2	& 44.8	& 3.7 \\
& Tango2	                &  14.1	& 51.7	& 3.7	& 17.4	& 62.2	& 3.6 \\
& \textbf{Average}        & \textbf{13.4}	& \textbf{47.9}	& \textbf{3.6}	& \textbf{16.5}	& \textbf{59.1}	& \textbf{3.6} \\
	
		\midrule
		\hline	
	\end{tabular}
		\end{adjustbox}
	\end{table}
	
		\begin{table}[ht]
		\centering
		\caption{Decoding speed in \gls{ai} and \gls{ra} configurations at \gls{qp}22 according to the sequence and number of threads.}
		\label{tb:sequence_ra_qp22}
		\begin{adjustbox}{max width=1\columnwidth}
		\begin{tabular}{c|c|c c c|c c c}
		\hline
					\midrule
		\multicolumn{2}{c|}{} & \multicolumn{6}{c}{FPS}  \\
		\cmidrule{3-8}	
		\multicolumn{2}{c|}{} &  \multicolumn{3}{c|}{AI} & \multicolumn{3}{c}{RA} \\
		\multicolumn{2}{c|}{} & 1 th. & 4 th. & Sp-up & 1 th. & 4 th. & Sp-up \\
					\midrule
		\multirow{6}{*}{HD} & FourPeople	        & 59.5	& 227.3	& 3.8	& 184.6	& 402.7	& 2.2  \\
& Johnny	        & 78.1	& 300.0	& 3.8	& 196.7	& 480.0	& 2.4  \\
& KristenAndSara	& 74.3	& 288.5	& 3.9	& 176.5	& 416.7	& 2.4  \\
& SlideEditing	        & 48.1	& 181.0	& 3.8	& 309.3	& 652.2	& 2.1  \\
& SlideShow	        & 108.6	& 420.0	& 3.9	& 241.5	& 617.3	& 2.6  \\
& \textbf{Average}	        & \textbf{73.7}	& \textbf{283.3}	& \textbf{3.8}	& \textbf{221.7}	& \textbf{513.8}	& \textbf{2.3}  \\
					\midrule				
		\multirow{7}{*}{FHD} & ArenaOfValor	        & 18.7	& 72.1	& 3.9	& 51.9	& 140.5	& 2.7  \\
& BasketballDrive	& 22.1	& 82.9	& 3.8	& 43.0	& 127.2	& 3.0  \\
& BQTerrace	        & 15.6	& 60.5	& 3.9	& 35.7	& 100.3	& 2.8  \\
& Cactus	        & 18.7	& 71.6	& 3.8	& 51.4	& 151.1	& 2.9  \\
& MarketPlace	        & 25.2	& 97.4	& 3.9	& 43.4	& 130.2	& 3.0  \\
& RitualDance	        & 31.3	& 117.2	& 3.8	& 49.4	& 153.5	& 3.1  \\
& \textbf{Average}	        & \textbf{21.9}	& \textbf{83.6}	& \textbf{3.8}	& \textbf{45.8}	& \textbf{133.8}	& \textbf{2.9}  \\
					\midrule
		\hline	
	\end{tabular}
		\end{adjustbox}
	\end{table}
	
			\begin{figure*}[ht]
		\begin{minipage}[b]{0.5\linewidth}
			\begin{subfigure}[b]{\linewidth}
				\centerline{\includegraphics[width=1\linewidth]{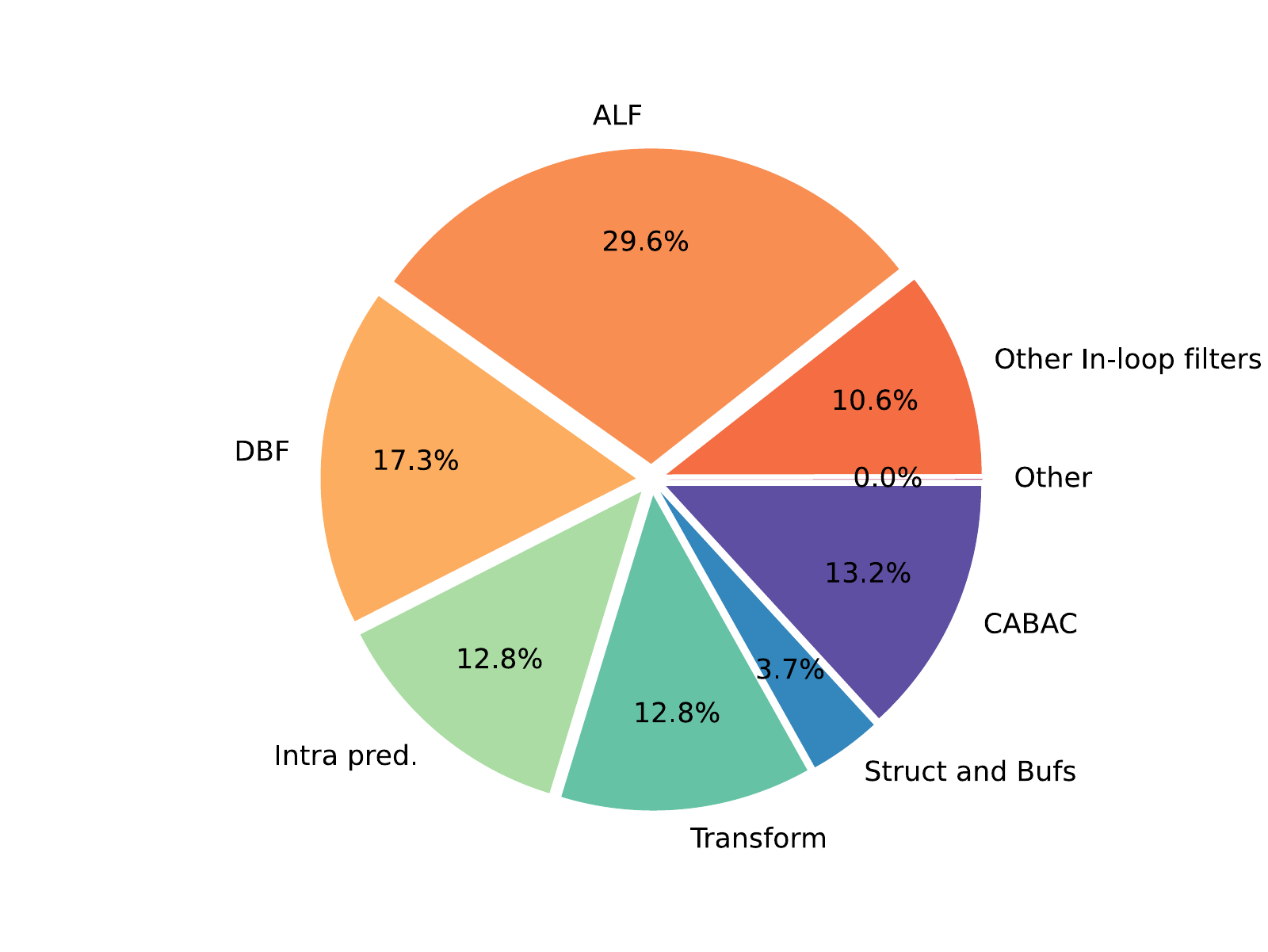}}
				\caption{QP=27.}
				\label{fig:profiling_ai_27}
			\end{subfigure}
		\end{minipage}
		\begin{minipage}[b]{0.5\linewidth}
			\begin{subfigure}[b]{\linewidth}
				\centerline{\includegraphics[width=1\linewidth]{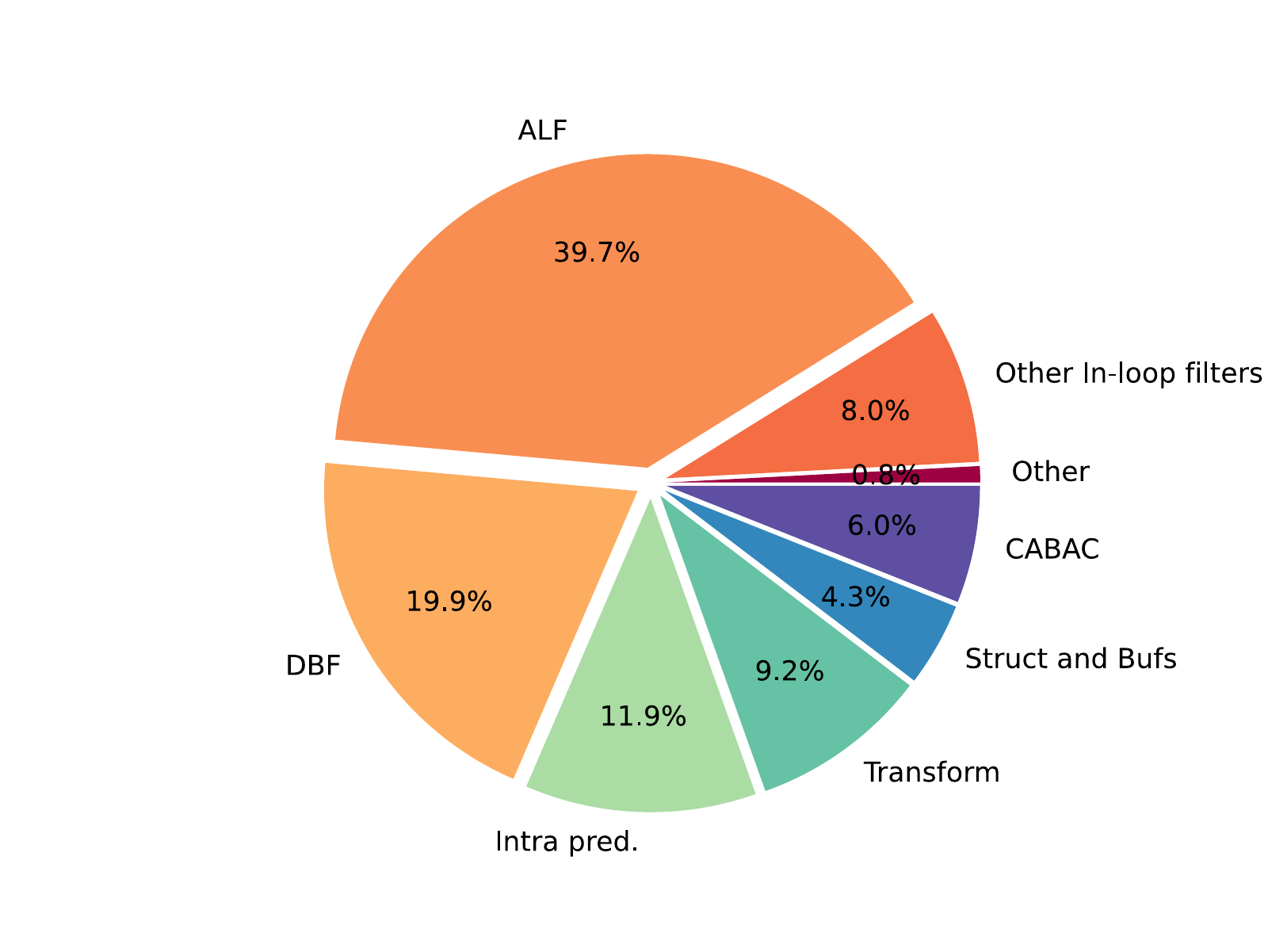}}
				\caption{QP=37.}
				\label{fig:profiling_ai_37}
			\end{subfigure}
		\end{minipage}
		\caption{Complexity distribution for the \gls{uhd} test sequence \emph{CatRobot1} in \gls{ai} configuration.}
		\label{fig:profiling_decoder_ai}
	\end{figure*}
	
	\Table{\ref{tb:sequence_ra}} presents the decoding speed obtained with \emph{OpenVVC} in \gls{ra} configuration according to the sequence and \gls{qp} value.	
	Many observations about the decoding speed disparities that were made in \gls{ai} configuration also apply in \gls{ra} configuration.
	Indeed, the decoding frame-rate increases with the \gls{qp} value and the sequences previously mentioned with more complex spatial textures lead lower decoding speed also in \gls{ra} configuration. 
	
	\Table{\ref{tb:sequence_ra}} points out the direct link between the speed-up obtained with 4 threads and the sequence resolution.
	Indeed, the speed-up variance is low among sequences with similar resolution, since it is contained in the short intervals $[2.2, 2.8]$, $[2.8, 3.3]$ and $[3.5, 3.7]$ for \gls{hd}, \gls{fhd} and \gls{uhd} resolutions, respectively.
	These numbers show that the speed-up is considerably impacted by the resolution of the sequence.
	As explained in Section~\ref{subsubsec:frame_par}, the \gls{mc} synchronization between decoding threads has been designed on a \gls{ctu} line basis. 
	For 128$\times$128 samples \glspl{ctu}, a \gls{ctu} line represents a 6th of a \gls{hd} picture against a 17th of a \gls{uhd} frame. 
	Therefore, at least a 6th of the \gls{hd} reference picture must be fully decoded before its data is used for \gls{mc}.
	The interactions among decoding threads for \gls{mc} synchronization will be higher, resulting in a lower speed-up compared to \gls{fhd} and \gls{uhd} resolutions. 
	
	Table~\ref{tb:sequence_ra_qp22} gives the decoding frame rate of {\it OpenVVC} in both \gls{ai} and \gls{ra} configurations at high bitrate (QP=22). In this specific configuration, the decoding frame rate of \gls{fhd} resolution with 4 threads is in average higher than 80 and 130 \gls{fps} for \gls{ai} and \gls{ra} configurations, respectively.

	\subsection{Complexity distribution in OpenVVC}
	
	The decoding complexity distribution is obtained by running \emph{OpenVVC} with Callgrind\footnote{Callgrind: \url{http://valgrind.org/docs/manual/cl-manual.html}.}.
	Callgrind is the Valgrind profiling tool that records the call history of program functions as a call graph.
	By default, the data collected includes the number of instructions executed, the calling/callee relation among functions and the number of calls.
	In contrast to execution time, that depends among others on memory accesses or \gls{cpu} frequency, the insight of the complexity distribution given by Callgrind is nearly constant regardless of the execution platform. 
	
	\subsection{AI coding configuration}
	

				\begin{figure*}[ht]
		\begin{minipage}[b]{0.5\linewidth}
			\begin{subfigure}[b]{\linewidth}
				\centerline{\includegraphics[width=1\linewidth]{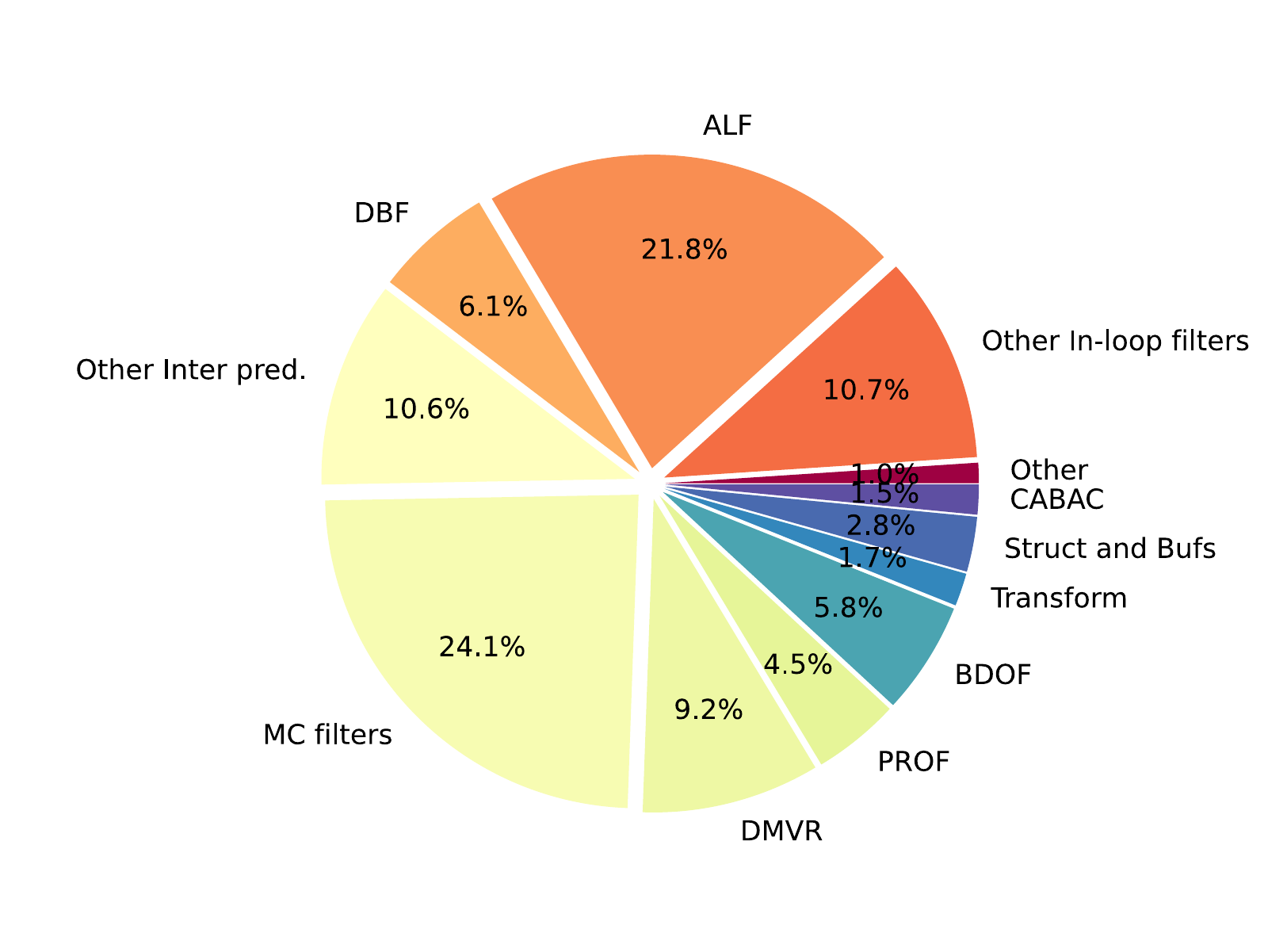}}
				\caption{QP=27.}
				\label{fig:profiling_ra_27}
			\end{subfigure}
		\end{minipage}
		\begin{minipage}[b]{0.5\linewidth}
			\begin{subfigure}[b]{\linewidth}
				\centerline{\includegraphics[width=1\linewidth]{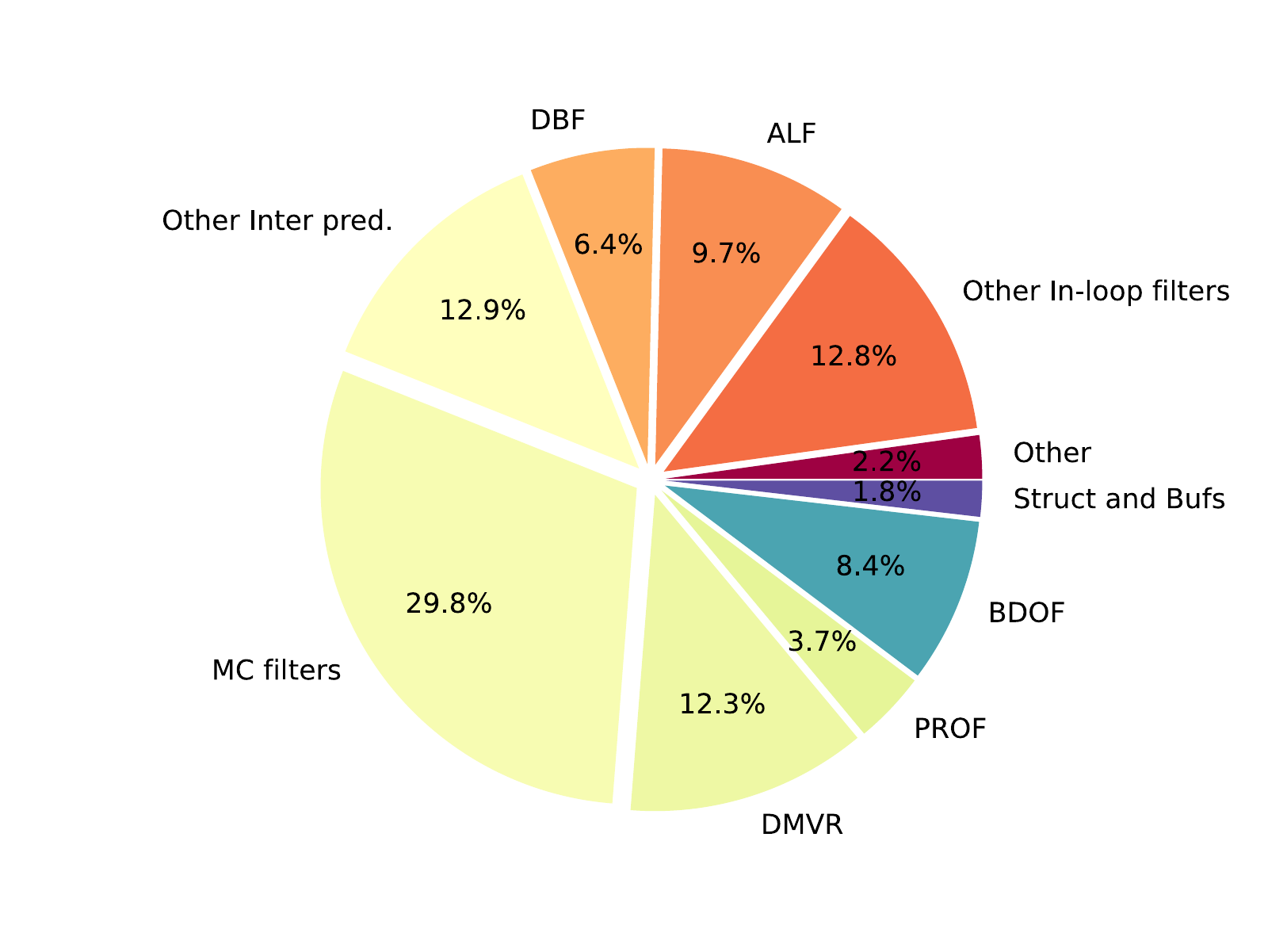}}
				\caption{QP=37.}
				\label{fig:profiling_ra_37}
			\end{subfigure}
		\end{minipage}
		\caption{Complexity distribution for the \gls{uhd} test sequence \emph{CatRobot1} in \gls{ra} configuration.}
		\label{fig:profiling_decoder_ra}
	\end{figure*}
	
	\Figure{\ref{fig:profiling_decoder_ai}} shows the decoding complexity distribution of \emph{OpenVVC} in \gls{ai} coding configuration, for \gls{uhd} test sequence \emph{CatRobot1} encoded at two \gls{qp} values  (\Figure{\ref{fig:profiling_ai_27}} for QP27  and \Figure{\ref{fig:profiling_ai_37}} for QP37).  
	The results are shown under the form of pie charts, in \% of total number of decoding instructions.
	The main decoding tasks displayed in \Figure{\ref{fig:profiling_decoder_ai}} have been presented in Section~\ref{section:overview_vvc_dec}.	
	The \gls{cabac} stage extracts from the bitstream the input data for all the other decoding stages. As mentioned in Section~\ref{subsec:simd}, the \gls{cabac} does not include significant data level parallelism and therefore is not accelerated with \gls{simd} optimizations.
	It explains its relatively high share of the decoding complexity at QP27 ($13.2$\%).
	For \gls{qp}37, the bitstream size is in average divided by 2.5 compared to \gls{qp}27.
	It results in a considerable decrease in the \gls{cabac} complexity ($6$\%) due to the lower amount of symbols to process.
	The intra prediction stage includes among others the application of angular, DC and Planar modes, as well as alternative \gls{waip}, \gls{mrl} and \gls{mip} modes.
	The intra prediction stage represents around 12\% of total complexity regardless the \gls{qp} value.
	The transform stage in \Figure{\ref{fig:profiling_decoder_ai}} computes the residual block through inverse quantization and inverse transform. 
	It also includes the aggregation of the predicted block with the residual block. 
	The transform stage is responsible for 12.8\% and 9.2\% of total complexity at \gls{qp}27 and \gls{qp}37, respectively. 
	This difference is due to the transform skip mode, which is more selected by the encoder at high \glspl{qp}.
	Four in-loop filters are performed on the reconstructed samples.
	They are displayed in shades of orange color in \Figure{\ref{fig:profiling_decoder_ai}}.
	The \gls{alf} provides a significant improvement in encoding efficiency~\cite{bross_overview_2021}.
	However, in counterpart of the aforementioned benefits, \gls{alf} represents a considerable burden for the decoding process (29\% and 39\% according to the \gls{qp} value). 
	The \gls{dbf} comes second with a share of over 17.3-19.9\% of total decoding complexity.
	In total, the in-loop filters are responsible for over 50\% of the decoding complexity in \gls{ai} configuration.  
	Finally, the operations on \emph{OpenVVC} structures and buffers, presented in Sections~\ref{subsec:frame_buffers} and~\ref{subsec:local_buffers}, represent 3.7-4.3\% of \emph{OpenVVC} complexity.

	\subsection{RA coding configuration}
	

	\Figure{\ref{fig:profiling_decoder_ra}} shows the decoding complexity distribution of \emph{OpenVVC} in \gls{ra} coding configuration, for \gls{uhd} test sequence \emph{CatRobot1} and according to the \gls{qp} value (\Figure{\ref{fig:profiling_ra_27}} for QP27 and \Figure{\ref{fig:profiling_ra_37}} for QP37).  
	{It is important to note the lower share of the complexity required by the \gls{alf} in \gls{ra} configuration (21.8\% and 9.7\%) compared to \gls{ai} configuration.
	Indeed, the \gls{alf} is disabled on a large number of \glspl{ctu} in \gls{ra} configuration, which is not the case in \gls{ai} configuration.}  
	In total, the sum of in-loop filters, \gls{cabac}, transform, intra prediction and internal buffers management represent 51\% and 41\% of the decoding complexity at QP27 and QP37, respectively.
	The remaining complexity is caused by the inter prediction tools, illustrated in shades of yellow color in the pie chart.
	The predominance of inter predicted frames in \gls{ra} configuration explains this number, and also explains the very low portion of intra prediction in the pie charts.
	In \gls{vvc} standard, the inter prediction stage enables various coding tools. 
	As shown in \Figure{\ref{fig:profiling_decoder_ra}}, the most complex tools are the \gls{mc} interpolation filters followed by \gls{dmvr}, \gls{bdof} and \gls{prof}.
	Theses thee last tools require the application of the \gls{mc} interpolation filters on the predicted block.
	For this reason, the complexity share of the \gls{mc} interpolation filters is higher than 24\% for both \gls{qp} values. 
	
	To summarize, this section has identified two decoding stages as complexity bottlenecks for \emph{OpenVVC}.
	The in-loop filtering stage is responsible for over 50\% of the decoding complexity in \gls{ai} configuration and the inter prediction stage is responsible for over 60\% of the decoding complexity in \gls{ra} configuration.
	In future works, the efforts for data and high level parallelism will therefore focus on these two decoding stages.

	\section {Conclusion}
	\label{sec:conc}
	
	In this paper, we presented \emph{openVVC}, an open source software  \gls{vvc} decoder that supports a broad range of \gls{vvc} tools.
	By combining extensive data level parallelism with frame level parallelism, \emph{OpenVVC} achieves real-time decoding for \gls{uhd} content.
	Considerable effort has been devoted to minimize both local and global buffer dimensions.
	As a consequence, the memory required by \emph{OpenVVC} is remarkably low, which is a great advantage for its integration on embedded platforms with low memory ressources.
	Compared to other \gls{sota} open source \gls{vvc} decoders, \emph{OpenVVC} achieves higher decoding speed than \emph{VVdeC} and reference software VTM in \gls{ai} configuration.
	In \gls{ra} configuration, the small gap with \emph{VVdeC} may be filled by implementing additional \gls{simd} optimizations and by combining frame level parallelism with other high level parallelism techniques, such as tile or wavefront.
	


	\def\url#1{}
	\bibliographystyle{IEEEbib}
	\bibliography{Bibliography}
	
\end{document}